\documentclass[fleqn,usenatbib]{mnras}

\usepackage{newtxtext,newtxmath}

\usepackage[T1]{fontenc}
\usepackage{xcolor}
\DeclareRobustCommand{\VAN}[3]{#2}
\let\VANthebibliography\thebibliography
\def\thebibliography{\DeclareRobustCommand{\VAN}[3]{##3}\VANthebibliography}

\usepackage{graphicx}	
\usepackage{amsmath}	
\usepackage{multirow}

\newcommand{\angstrom}{\mbox{\normalfont\AA}}
\usepackage{xspace}
\newcommand{\ergscmA}{\hbox{$\mathrm{erg\,s^{-1}\,cm^{-2}\,\angstrom^{-1}}$\xspace}}

\title[IGM Surrounding GRB\,210905A]{Neutral Fraction of Hydrogen in the Intergalactic Medium Surrounding High-Redshift Gamma-Ray Burst\,210905A}

\author[H. M. Fausey et al.]{H.M. Fausey,$^{1}$\thanks{E-mail: hfausey@gwu.edu}
S. Vejlgaard,$^{2,3}$
A.J. van der Horst,$^{1}$
K.E. Heintz,$^{2,3}$
L. Izzo,$^{4}$
D.B. Malesani,$^{5,2,3}$ \newauthor
K. Wiersema,$^{6,7}$
J.P.U. Fynbo,$^{2,3}$
N.R. Tanvir,$^{8}$
S.D. Vergani,$^{9,10,11}$
A. Saccardi,$^{9}$
A. Rossi,$^{12}$\newauthor
S. Campana,$^{10}$
S. Covino,$^{10}$
V. D'Elia,$^{13, 14}$
M. De Pasquale$^{15}$
D. Hartmann,$^{16}$
P. Jakobsson,$^{17}$\newauthor
C. Kouveliotou,$^{1}$
A. Levan,$^{5,18}$
A. Martin-Carrillo,$^{19}$
A. Melandri,$^{14}$
J. Palmerio,$^{9,11}$
G. Pugliese,$^{20}$ \newauthor
R. Salvaterra$^{21}$
\\
$^{1}$Department of Physics, George Washington University, 725 21st St. NW, Washington, DC, 20052, USA\\
$^{2}$Cosmic Dawn Center (DAWN), Copenhagen, Denmark\\
$^{3}$Niels Bohr Institute, University of Copenhagen, Jagtvej 128, 2200 Copenhagen N, Denmark\\
$^{4}$DARK, Niels Bohr Institute, University of Copenhagen, Jagtvej 128, 2200 Copenhagen, Denmark\\
$^{5}$Department of Astrophysics/IMAPP, Radboud University, 6525, AJ Nijmegen, The Netherlands\\
$^{6}$Department of Physics, Astronomy and Mathematics, University of Hertfordshire, Hertfordshire AL10 9AB, UK\\
$^{7}$Physics Department, Lancaster University, Lancaster, LA1 4YB, UK\\
$^{8}$School of Physics and Astronomy, University of Leicester, University Road, Leicester LE1 7RH, UK\\
$^{9}$GEPI, Observatoire de Paris, Universit\'e PSL, CNRS, 5 pace Jules Janssen, 92190 Meudon, France\\
$^{10}$INAF -- Osservatorio Astronomico di Brera, Via E. Bianchi 46, 23807 Merate, LC, Italy\\
$^{11}$Institut d'Astrophysique de Paris, UMR 7095, CNRS-SU, 98 bis boulevard Arago, 75014 Paris, France\\
$^{12}$INAF -- Osservatorio di Astrofisica e Scienza dello Spazio, Via Piero Gobetti 93/3, 40129 Bologna, Italy\\
$^{13}$Space Science Data Center (SSDC) -- Agenzia Spaziale Italiana (ASI), 00133 Roma, Italy\\
$^{14}$INAF -- Osservatorio Astronomico di Roma, Via Frascati 33, 00040 Monte Porzio Catone, Italy\\
$^{15}$MIFT Department, Polo Papardo, University of Messina, Via Ferdinando Stagno d'Alcontres 31, 98166 Messina, Italy\\
$^{16}$Department of Physics \& Astronomy, Clemson University, Kinard Lab of Physics, Clemson, SC 29634, USA\\
$^{17}$Center for Astrophysics and Cosmology, Science Institute, University of Iceland, Dunhagi 5, 107 Reykijav\'{i}k, Iceland\\
$^{18}$Department of Physics, University of Warwick, Coventry, CV4 7AL, UK\\
$^{19}$ School of Physics and Centre for Space Research, University College Dublin, Belfield, Dublin 4, Ireland\\
$^{20}$Astronomical Institute Anton Pannekoek, University of Amsterdam, 1090 GE Amsterdam, The Netherlands\\
$^{21}$INAF -- Istituto di Astrofisica Sapziale e Fisica Cosmica, Via Alfonso Corti 12, 20133, Milano, Italy
}

\date{Accepted XXX. Received YYY; in original form ZZZ}

\pubyear{2023}

\begin{document}
\label{firstpage}
\pagerange{\pageref{firstpage}--\pageref{lastpage}}
\maketitle

\begin{abstract}
The Epoch of Reionization (EoR) is a key period of cosmological history in which the intergalactic medium (IGM) underwent a major phase change from being neutral to almost completely ionized.
Gamma-ray bursts (GRBs) are luminous and unique probes of their environments that can be used to study the timeline for the progression of the EoR.
Here we present a detailed analysis of the ESO Very Large Telescope X-shooter spectrum of GRB\,210905A, which resides at a redshift of $z\sim 6.3$.
We focus on estimating the fraction of neutral hydrogen, $x_{\rm H_I}$, on the line of sight to the host galaxy of GRB\,210905A by fitting the shape of the Lyman-$\alpha$ damping wing of the afterglow spectrum.
The X-shooter spectrum has a high signal to noise ratio, but the complex velocity structure of the host galaxy limits the precision of our conclusions.
The statistically preferred model suggests a low neutral fraction with a 3-$\sigma$ upper limit of $x_{\rm H_I} \lesssim 0.15$ or $x_{\rm H_I} \lesssim 0.23$, depending on the absence or presence of an ionized bubble around the GRB host galaxy, indicating that the IGM around the GRB host galaxy is mostly ionized.
We discuss complications in current analyses and potential avenues for future studies of the progression of the EoR and its evolution with redshift.

\end{abstract}




\section{Introduction}

Gamma-ray bursts (GRBs) are extremely bright transient events that can be observed from across the Universe.
GRBs have two phases: the prompt emission, in which internal shocks due to variability in the outflow \citep{Rees1992} or magnetic reconnection \citep{Thompson1994, Spruit2001, Giannios2007, Lyubarsky2010, Beniamini2016} produce a flash of gamma rays; and the afterglow, multi-wavelength emission due to the relativistic jet interacting with the surrounding medium \citep{Sari1998}.

GRBs can be divided into two classes: short-hard and long-soft GRBs \citep{Mazets1981, Kouveliotou1993}.
Short-hard GRBs are thought to arise from compact object mergers \citep{Eichler1989, Narayan1992}. 
They have a prompt emission that usually lasts less than 2 seconds, and tend to have harder spectra \citep{Kouveliotou1993}.
Long-soft GRBs are usually associated with the core collapse of Wolf-Rayet stars \citep{Woosley1993, Galama1998, Chevalier1999, Hjorth2003}.
Their prompt emission generally lasts longer than 2 seconds and they tend to have softer spectra \citep{Kouveliotou1993}.
However, recently some long-duration GRBs with kilonova counterparts have been associated with compact object mergers \citep{Gao2022, Rastinejad2022, Levan2024}, and a short-duration GRB with a soft spectrum and associated supernova \citep{Rossi2022b} have been observed.

While both classes of GRBs are extremely luminous, long GRBs can be particularly bright, with some having isotropic equivalent luminosities larger than $10^{54}~\text{erg~s}^{-1}$ \citep{Frederiks2013, Burns2023}.
This allows long GRBs to potentially be detected out to $z\sim 20$ \citep{Lamb2000}, making them cosmological probes of the high-redshift Universe \citep{Campana2022}.
The afterglows of GRBs have a simple power-law spectrum \citep[e.g.,][]{Sari1998} as compared to other high-redshift probes such as quasars, Lyman-$\alpha$ emitters, and Lyman Break Galaxies, which have a relatively complex continuum spectra.
GRBs are also short-lived compared to quasars, which emit a steady stream of ionizing radiation into the surrounding medium, potentially biasing them towards higher ionization states \citep{Totani2006}.
Because GRBs do not emit ionizing radiation for an extended period of time, there is no need to correct for additional IGM ionization from the GRB.
For these reasons, GRBs are ideal for studying the chemical evolution of galaxies \citep{Savaglio2006, Thone2013, Sparre2014, Saccardi2023, Heintz2023} and the Epoch of Reionization \citep[EoR;][]{MiraldaEscude1998, Totani2006, McQuinn2008, Hartoog2015, Lidz2021}, and could potentially contribute to understanding early star formation and the initial mass function \citep{Lloyd2002, Fryer2022} as well as population III stars \citep{Lloyd2002, Campisi2011}.

The EoR is a key era in cosmological history, yet is still poorly understood.
It was likely driven by ionizing radiation from the first galaxies, with possible contributions from faint active galactic nuclei \citep{Arons1972, Tegmark1994, Haiman1996, MiraldaEscude1998, Madau1999, Faucher2008, Bouwens2012, Becker2013, Giallongo2015, Chardin2015, Madau2015, McQuinn2016, Matthee2023}. 
Models and observations suggest that the EoR ended around $z\sim 5.5 - 6$ \citep{Totani2006, Robertson2015, Ishigaki2018, Finkelstein2019, Naidu2020, Qin2021}, but its progression at higher redshifts is poorly understood. Quasars have been used to track the progression of the EoR by estimating the fraction of neutral hydrogen, i.e., the neutral fraction $x_{\rm H_I}$, in the intergalactic medium (IGM) along the line of sight \citep{Mortlock2011, Simcoe2012, Bosman2015, Greig2017, Banados2018, Davies2018, Durovcikova2020, Wang2020, Yang2020, Greig2022, Fan2023}.
There have been fewer than 10 GRBs with spectroscopic or photometric redshifts greater than $6$ to date \citep{Tanvir2009, Cucchiara2011, Salvaterra2015, Tanvir2018}, and a few of these have been used as cosmological probes of the EoR.
The latter is done by examining the shape of the red Lyman-$\alpha$ (Ly$\alpha$) damping wing, which provides insight into the ionization state of the IGM surrounding the host galaxy by estimating the fraction of neutral hydrogen \citep{MiraldaEscude1998, Totani2006, Totani2014, Hartoog2015, Totani2016}.

GRB\,210905A was a $z \sim 6.3$ GRB with one of the most luminous late-time optical afterglows ever observed, which allowed for detailed multi-wavelength analyses \citep{Rossi2022}. 
Its redshift was determined using metal and fine structure lines associated with the GRB host galaxy \citep{GCN, Rossi2022, Saccardi2023}.
Spectroscopy with the X-shooter instrument on the ESO Very Large Telescope (VLT) showed two absorption systems around $z\,=\,6.3118$ and $z\,=\,6.3186$, as well as additional intervening absorbers at $z\,=\,5.739$ and $z\,=\,2.830$ \citep{Saccardi2023}.
The two velocity systems at $z\approx 6.3$, with a velocity difference of $278~\text{km s}^{-1}$ and both associated with the host galaxy, were well identified in the metal lines, allowing for an analysis of their metallicity and chemical composition \citep{Saccardi2023}.
The metalicity of the $z\sim 6.3$ system was found to be $\left[\rm{M/H}\right]_{\rm tot} = -1.72 \pm 0.13$ \citep{Saccardi2023} which is consistent with observations for other GRB DLAs at $z \sim 6$, and is on the higher end of metallicities for quasar DLAs when extrapolated to high redshift \citep{DeCia2018, Saccardi2023}.

Determining the neutral fraction of the IGM at different redshifts is important to estimate the progression of the EoR.
In this paper, we use the X-shooter spectrum of GRB\,210905A to decouple the interstellar medium (ISM) and IGM contributions to the shape of the red Ly$\alpha$ damping wing, and obtain an estimate of the neutral fraction in the IGM surrounding its $z\sim 6.3$ host galaxy, using the model developed by \citet{MiraldaEscude1998}.
We perform detailed modeling using the multiple velocity systems found by \citet{Saccardi2023}, taking them to be damped Ly$\alpha$ systems DLAs; systems of gas and dust with a large amount of Ly$\alpha$ absorption due to neutral hydrogen) for this analysis. We also include considerations of the \citet{McQuinn2008} damping wing model to investigate the robustness of the result.
We use $H_0 = 67.4 ~\text{km/s/Mpc}$, $\Omega_m = 0.315$, $\Omega_{b}h^2 = 0.0224$, and $Y_\text{P} = 0.2454$ for the cosmological parameters throughout this paper \citep{Planck2020}.
In Section 2, we present the spectrum and models used in this study.
In Section 3, we present the results of our damping wing analysis for different assumptions regarding the DLA systems and spectral index. 
In Section 4, we use robust statistical methods to reach conclusions regarding the model parameters, in particular the neutral fraction. 
In Sections 5 and 6, we discuss the implications of the results for future high-redshift GRB analyses, and summarize our findings.

\section{Observations and Modeling}

\subsection{Data Reduction and Selection}
GRB\,210905A was observed with the X-shooter spectrograph \citep{Vernet2011}, mounted on the UT3 (Melipal) telescope, which is located at the VLT observatory in Cerro Paranal (Chile). Data acquisition started 2.53 hours after the GRB detection, and consisted of four nod-on-slit exposures of 20 minutes in the three arms of the instrument (UVB, VIS and NIR with K-band blocking filter), which cover the wavelength range from 3,000 to 21,000 \AA~ \citep{GCN}. The dataset was reduced using standard \textit{esorex} prescriptions \citep{Modigliani2010}, although for the UVB and VIS arms each single exposure was first reduced in STARE mode, after which all exposures were aligned and subsequently stacked to obtain the final 2D spectrum. 
Post-processing scripts described in \citet{Selsing2019} were applied to correct the final spectrum for slit losses, improve the residual sky lines correction, and extract the final 1D spectrum. 
The VIS arm was further corrected for telluric lines using a python wrapper for the line-by-line radiative transfer model \citep{Clough2014}, which provides a best-fit telluric spectrum obtained from a model of the atmospheric conditions at Paranal and a fit of the telluric lines in the observed dataset.
This spectrum is the same as those used in the \citet{Rossi2022} and \citet{Saccardi2023} analyses.

\begin{figure*}
    \centering
    \includegraphics[width = \linewidth]{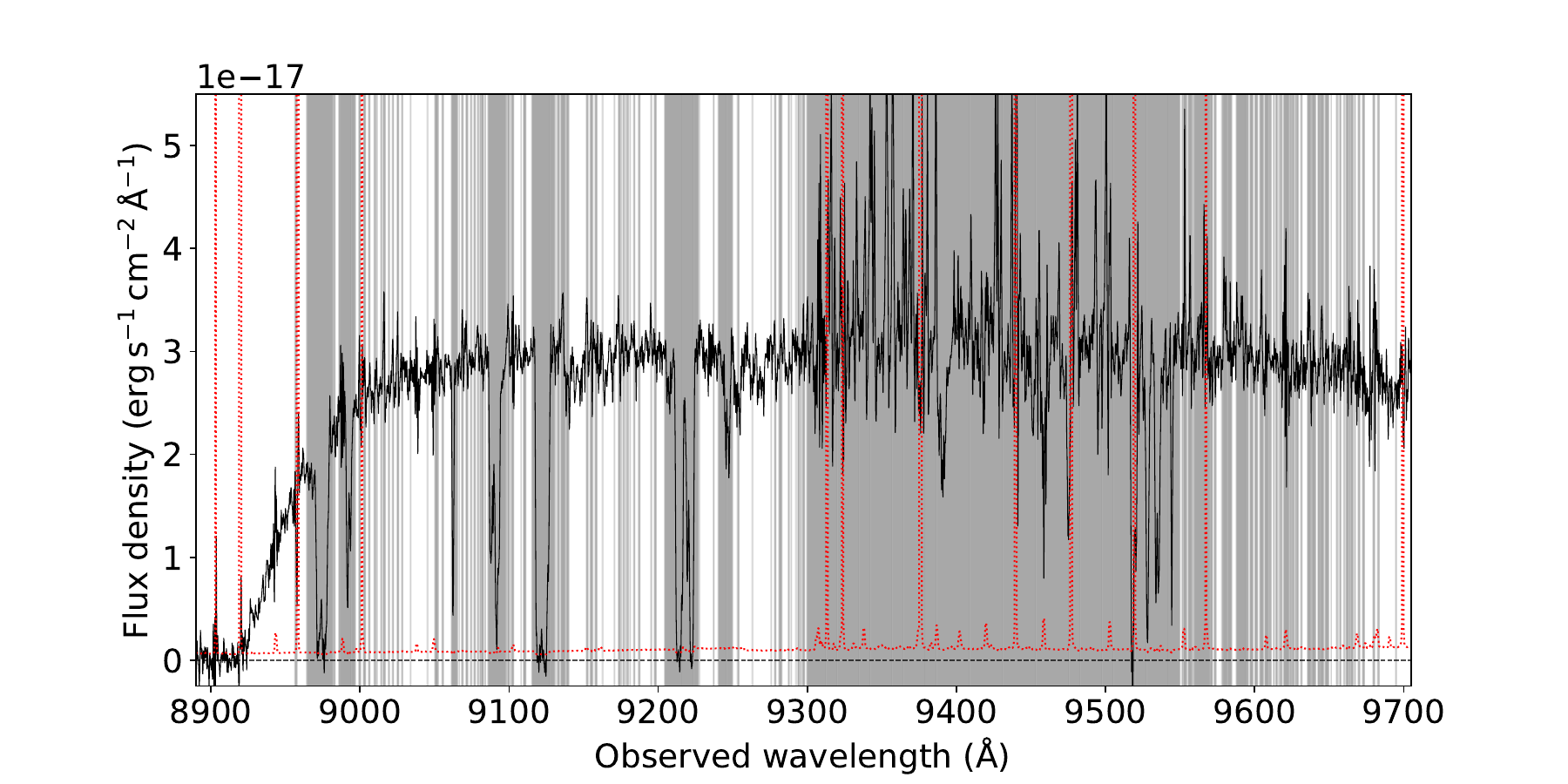}
    \caption{GRB\,210905A spectrum used for the Ly$\alpha$ damping wing analysis (solid black) with error spectrum (dotted red). Regions with metal or telluric lines are shaded in grey and are excluded from the analysis.}
    \label{fig:finalSpectrum}
\end{figure*}

For the damping wing analysis we chose to use only the VIS data, which ranges from 5,595 - 10,240 \AA~ \citep{Vernet2011} with 0.2~\AA~bins, to avoid any issues related to calibration offsets between the X-Shooter VIS and NIR arms. 
Based on the GRB redshift, the bottom of the red wing of the Ly$\alpha$ forest ends at 8,896~\AA~(i.e., $\lambda_\alpha(1+z)$ where $\lambda_\alpha = 1,215.67$~\AA~is the Ly$\alpha$ wavelength in a vacuum), and we focus on analyzing the spectrum redward of this wavelength.
We removed any wavelength bins with telluric absorption greater than 0.9 in either the \emph{esorex} prescriptions or the ESO Skycalc Model Calculator \citep{Noll2012, Jones2013}, as well as the metal absorption lines identified in \citet{Saccardi2023}.
We also excluded data between 9,300 - 9,530~\AA~due to the high density of absorption and emission lines in those regions. 
Finally, for our modeling we only used the data up to 9,700~\AA, which is the transition between the two largest-wavelength orders in the VIS arm, to avoid systematic errors that arise from transitions between these highly curved orders. The final spectrum used for the damping wing analysis is shown in Figure~\ref{fig:finalSpectrum}.
We note that there is still a deviation in the flux of this GRB compared to the error spectrum, which results in a relatively high chi-squared per degrees of freedom in our fitting.
For the data reduction we have followed the ESO recommended strategies, which are informed by processing many years of data in quality control.
We are confident that the data was processed in a way that minimizes any systematic behavior in error estimates, but any fit bias due to inaccuracy in the noise model cannot be completely excluded.

\section{Model and Fitting Methodology}

In this section, we describe the different models and fitting methodologies used to fit the Ly$\alpha$ damping wing.

\subsection{Model}
\label{sec:model}

\subsubsection{ISM contribution}

For the ISM contribution to the damping wing, we consider the two velocity systems identified at $z_{\rm DLA}=6.3118$ and $z_{\rm DLA} = 6.3186$.
\citet{Saccardi2023} take the GRB redshift to be $z=6.3118$, and conclude that the two velocity systems are likely  both from the ISM of the GRB host galaxy.
To ensure that we account for any potential absorption from each velocity system, for this analysis we will consider the two main absorption systems identified by \citet{Saccardi2023} as two DLAs. 
We adopt the model used in \citet{Totani2006} for DLA contributions to the damping wing:

\begin{equation}
    \tau_{\rm DLA}(\nu_{\rm obs}) = N_{\rm H_I}\sigma_\alpha\left[\nu_{\rm obs}(1 + z_{\rm DLA})\right]
\end{equation}
where $\nu_{\rm obs} = c/\lambda_{\rm obs}$, and $\sigma_\alpha(\nu)$ is defined as

\begin{equation}
    \sigma_\alpha(\nu) = \frac{3 \lambda_\alpha^2 f_\alpha \Lambda_{\rm cl,\alpha}}{2\pi} * \frac{\Lambda_\alpha \left(\frac{\nu}{\nu_\alpha}\right)^{4}}{16\pi^2(\nu - \nu_\alpha)^2 + \Lambda_\alpha^2\left(\frac{\nu}{\nu_\alpha}\right)^6}
\end{equation}

\noindent The rate of spontaneous radiative decay between levels in the Ly$\alpha$ transition is given by $\Lambda_\alpha = 3\left(g_u/g_l\right)^{-1}f_\alpha \Lambda_{\rm cl,\alpha}$, where $g_u/g_l = 3$ and $f_\alpha = 0.4162$ are the ratio of the weights of the upper and lower atomic levels and the absorption oscillator strength for the first Ly$\alpha$ transition, respectively \citep{Weisskopf1933, Peebles1993, Madau2000, Morton2003,Bach2014}, and $\Lambda_{\rm cl,\alpha} = 8\pi^2 e^2/3 m_e c \lambda_\alpha^2$ is the classical damping constant \citep{Weisskopf1933, Totani2006, Bach2014}. 

\subsubsection{\citet{MiraldaEscude1998} IGM model}

One model for the red Ly$\alpha$ damping wing is based on the one developed by \citet{MiraldaEscude1998} and used in previous analyses of other high-redshift GRB damping wings \citep{Totani2006, Chornock2013, Totani2014, Hartoog2015, Totani2016}.
It has also been used to estimate the neutral fraction in other fields \citep[e.g.,][]{Banados2018, Andika2023, Umeda2023}.
The model assumes a uniform distribution of neutral hydrogen between a fixed upper and lower redshift, $z_{\rm IGM,u}$ and $z_{\rm IGM,l}$. There is assumed to be no neutral hydrogen below $z_{\rm IGM,l}$. 
The presence of neutral hydrogen increases the optical depth of the medium surrounding the host galaxy, and alters the shape of the red damping wing, which allows the neutral fraction to be estimated.
The \citet{MiraldaEscude1998} model does not account for patchiness in the IGM. However, \citet{Chen2024} and \citet{Keating2024} find that the \citet{MiraldaEscude1998} model does sufficiently well at estimating the neutral fraction from the damping wing profile when compared to more complex simulations that account for a patchy Universe using the CROC \citep{CROC} and Sherwood-Relics \citep{Bolton2017, Puchwein2023} simulations, respectively. They argue that, while patchiness can introduce some scatter, the shape of the damping wing shape is most strongly affected by the average neutral hydrogen density along the line of sight. We also implement the \citet{MiraldaEscude1998} model in multiple shells with independent neutral fractions to examine the neutral fraction results in different redshift ranges (see Section \ref{sec:shells}). 

In the model developed by \citet{MiraldaEscude1998} and further refined by \citet{Totani2006}, the IGM  optical depth as a function of observed wavelength is described as

\begin{equation}
    \begin{aligned}
    \tau_{\rm IGM}(\lambda_{\rm obs}) = &\frac{x_{\rm H_{I}}\Lambda_\alpha \lambda_\alpha \tau_{GP}(z_{\rm host})}{4 \pi^2 c}\left[\frac{1 + z_{\rm obs}}{1 + z_{\rm host}}\right]^{3/2}
    \\
    & \times \left[I\left(\frac{1 + z_{\rm IGM,u}}{1 + z_{\rm obs}}\right) - I\left(\frac{1 + z_{\rm IGM,l}}{1 + z_{\rm obs}}\right)\right]
    \end{aligned}
\end{equation}

\noindent where $x_{\rm H_{I}}$ is the neutral fraction, $z_{\rm host}$ is the host galaxy redshift,$z_{\rm obs}$ is defined by $1 + z_{\rm obs} \equiv \lambda_{\rm obs}/\lambda_\alpha$, and $z_{\rm IGM,u}$ and $z_{\rm IGM,l}$ are the upper and lower redshifts between which the distribution of neutral hydrogen is assumed to be constant. Outside of this redshift range the neutral fraction is assumed to be negligible.
\citet{Totani2006} model the Gunn-Peterson optical depth using
\begin{equation}
    \tau_{\rm GP}(z) = \frac{3 f_\alpha \Lambda_{\rm cl,\alpha}\lambda_\alpha^3 \rho_{\rm crit}\Omega_{\rm B}(1-Y_{\rm p})}{8\pi m_p H_0 \Omega_{\rm M}^{1/2}}(1+z)^{3/2}
\end{equation} 
where $\rho_{\rm crit} = 3H_0^2/8\pi G$. The function $I(x)$ is defined as \citep{MiraldaEscude1998, Totani2006}

\begin{equation}
    I(x) = \frac{x^{9/2}}{(1-x)} + \frac{9}{7} x^{7/2} + \frac{9}{5} x^{5/2} + 3x^{3/2} + 9x^{1/2} - \frac{9}{2} \ln\left(\frac{1 + x^{1/2}}{1 - x^{1/2}}\right)
\end{equation}

We set $z_{\rm IGM,u}$ to the lower DLA redshift $z_{\rm IGM,u} = 6.3118$, since it is the lowest-redshift system that is likely to be part of the GRB host galaxy, and $z_{\rm IGM,l} = 6$ since this is the redshift at which the EoR is theorized to have ended \citep{Robertson2015, Ishigaki2018, Naidu2020}. 
Some models suggest that the Epoch of Reionization could have ended at $z\sim 5.5$ \citep[e.g.,][]{Finkelstein2019, Qin2021, Bosman2022}. However, the choice of $z_l$ does not have an effect on the neutral fraction result so long as it is assumed to be sufficiently far from the GRB redshift (see Section \ref{sec:zlchoice} for further discussion on the effect of the $z_l$ parameter).

\subsubsection{\citet{McQuinn2008} IGM model}

To assess the robustness of our results, we also performed fits using the \citet{McQuinn2008}  model, which accounts for patchiness in the IGM by assuming a comoving bubble of ionized hydrogen around the host galaxy with radius $R_{\rm b}$. 
This model is an approximate version of the \citet{MiraldaEscude1998} model, but allows the radius of an ionized bubble to be fit as a free parameter rather than assumed from the value of $z_{\rm IGM,u}$.
The \citet{McQuinn2008} model optical depth due to neutral hydrogen in the IGM as:

\begin{equation}
    \begin{aligned}
    \tau_{\rm IGM} \approx &900 \text{km s}^{-1} \times ~ x_{\rm H_I}\left(\frac{1+z_{\rm DLA}}{8}\right)^{\frac{3}{2}} 
    \\
    & \times \left[\frac{H(z_{\rm DLA}) R_{\rm b}}{1+z_{\rm DLA}} - c \left(\frac{\nu_{\rm obs}(1 + z) - \nu_\alpha}{\nu_\alpha}\right)\right]^{-1}
    \end{aligned}
\end{equation}

\noindent where $H(z) = H_0 \sqrt{\Omega_{\rm M}(1+z)^3 + \Omega_\Lambda + \Omega_k(1+z)^2}$ \citep{Farooq2013, Verde2014, Chen2017, Wei2017}. 
When fitting with the \citet{McQuinn2008} IGM model, we use the same parameters as for the \citet{MiraldaEscude1998} model, but without $z_{\rm IGM,u}$ and $z_{\rm IGM,l}$, and with the additional parameter $R_{\rm b}$. 
This model does not account for additional patchiness outside of the ionized region directly around the host galaxy. However, \citet{McQuinn2008} test their model against three radiative transfer simulations and find that the resulting $R_{\rm b}$ estimates always roughly match the actual $R_{\rm b}$ value, and the measured $x_{\rm H_I}$ along the line of sight is always within 0.3 of the global neutral fraction value.

\subsection{Fitting}

For the fitting, we use four parameters: the normalization $A$ at 10,000~\AA; the continuum spectrum power law index $\beta$; the column density $N_{\rm H_I}$ for the DLAs; and our main parameter of interest, the neutral fraction $x_{\rm H_I}$.
We do not fit for extinction since previous analyses found dust extinction to be negligible for GRB\,210905A \citep{Saccardi2023}.
The column density and neutral fraction are not independent of one another, so we perform a joint fit of the DLA column densities and neutral fraction in an attempt to disentangle their contributions to the damping wing. 
Because there are two DLAs, we use two column density parameters, one for each DLA (but we also show one-DLA fits for comparison). 
The column densities of different elements in the two systems are very similar, so we perform some fits of the data with a fixed difference between their H$_{\rm I}$ column densities according to $\log(N_{\rm H_I,6.3118}) = \log(N_{\rm H_I,6.3186}) + 0.38$, based on the difference of their average Si$_{\rm II}$ column densities, $14.36$ and $14.74$ at $z = 6.3186$ and $z = 6.3118$, respectively, as found by \citet{Saccardi2023}.
We also perform modeling with the column densities of the two DLAs uncoupled to see the the impact on the results when attempting an independent measurement of the two DLA column densities.

We are only fitting the VIS data to avoid any offsets between the VIS and NIR arms, and we model the data up to 9,700~\AA, as discussed in the previous subsection. While this approach results in a more reliable estimate of our key parameters of interest, namely $N_{\rm H_I}$ and $x_{\rm H_I}$, it also means that there is not enough of a lever arm to confidently constrain the spectral index. We fix the value of the spectral index to results reported in the literature from spectral energy distribution (SED) fits to multi-wavelength data, but include fits with the spectral index as a free parameter as well for completeness.

There are conflicting results for the optical spectral index based on X-ray to NIR analysis. 
The \emph{Swift}-XRT spectrum repository reports a late-time X-ray photon index $\Gamma$ of $1.90 \pm 0.15$ \citep[with 90\% uncertainties;][]{Evans2009}, where $n(E)dE \propto E^{-\Gamma}$.
This photon index corresponds to an X-ray spectral index $\beta_X$ of $0.90 \pm 0.09$ (with 1-$\sigma$ uncertainties). 
Multi-wavelength analysis reveals that the SED is consistent with slow cooling, with the synchrotron cooling break falling between the X-ray and optical regimes of the spectrum \citep{Rossi2022}.
Assuming that this break is the cooling break, $\beta_{\text{opt}} = \beta_{X} - 1/2$ leads to an optical spectral index of $0.40 \pm 0.09$ \citep{Sari1998}. 
In their X-ray to NIR analysis, \citet{Rossi2022} find a spectral index (with 1-$\sigma$ uncertainty) of $0.60\pm 0.04$. 
We perform fits fixing the spectral index to both $\beta = 0.40$ and $\beta = 0.60$ to account for both possible spectral indices in our modeling (see Table \ref{tab:results}).
A similar approach for the spectral index has been taken for other GRB damping wing analyses \citep[e.g.,][]{Hartoog2015}. We note that a spectral fit of the X-shooter NIR arm returns a spectral index of $\sim 0.45$, but we note that the NIR arm is much noisier than the VIS arm in X-shooter.

The spectral fitting is performed using the \texttt{emcee} python package implementation of the Markov-Chain Monte-Carlo method \citep[MCMC;][]{emcee}.
MCMC methods estimate parameter values by constructing posterior distributions by combining a likelihood function with a set of priors, and probing the resulting parameter space using a certain number of walkers \citep{MCMCtextbook}.
To fit the Ly$\alpha$ damping wing using the \citet{MiraldaEscude1998} model, we use 50 walkers with a 1500 step burn-in and 3000 step production chain, which is sufficiently long for the walkers to settle into a stable region of parameter space. 
We increase the length of the burnin and production chains (5000 and 10000 steps, respectively) for the \citet{McQuinn2008} model to adjust for the additional parameters.
We use a Bayesian likelihood function $\log(\mathcal{L}) = -\chi^2/2$, where $\mathcal{L}$ is the likelihood and $\chi^2$ is the standard definition of chi-squared.
All parameters have linearly uniform priors, but we restrict the normalization to $A > 0$, the DLA column density to $18< \log(N_{\rm H_I}/\text{cm}^{-2}) < 23$ \citep{Tanvir2019}, and the neutral fraction to $0 \leq x_{\rm H_I} \leq 1$. For fits in which $\beta$ is a free parameter, we use a linearly uniform prior restricting the spectral index to $0 \leq \beta \leq 2$.
For fits using the \citet{McQuinn2008} model, we use a linearly uniform prior for $R_{\rm b}$, but restrict the bubble size to $0~\text{Mpc}h^{-1} \leq R_{\rm b} \leq 60~\text{Mpc}h^{-1}$ (or $\sim 90~\text{Mpc}$), since the latter corresponds to the \citet{Lidz2021} prediction for ionized bubble size for a largely ionized ($x_{\rm H_I} \sim 0.05)$ IGM. 
We focus on fits using this prior, but also include results with $R_{\rm b} \leq 130~\text{Mpc}\;h^{-1}$ (or $\sim 193~\text{Mpc}$, corresponding to $z \sim 6.0$) and $R_{\rm b} \leq 355~\text{Mpc}\;h^{-1}$ (or $527~\text{Mpc}$, corresponding to $z \sim 5.5$) to ensure that our choice of prior is not biasing the results (see Appendix \ref{sec:additionalMcquinn}).

Fitting results using the \citet{MiraldaEscude1998} model are reported in Sections \ref{sec:results} and \ref{sec:compare}; fitting results using the \citet{McQuinn2008} are reported in Section \ref{sec:McQuinn}; and fitting results from the \citet{MiraldaEscude1998} shell implementation are discussed in Section \ref{sec:shells}.

\section{Results}
\label{sec:results}

To measure the fraction of neutral hydrogen in the IGM surrounding the GRB host galaxy, we perform a fit of the Ly$\alpha$ damping wing using spectroscopic data from VLT/X-shooter instrument.
As described in the previous section, we perform multiple fits with $\beta$ fixed to two different values, as well as an additional free $\beta$ fit. We consider both coupled and uncoupled column densities for the two velocity systems in the GRB host galaxy; and explore how the results change when we fit the damping wing with 
and we include fits for one DLA at the GRB redshift for completeness and comparison.
The results for the two-DLA fits are presented in Table \ref{tab:results} and discussed in the following subsections. The fits for one DLA are discussed in Section~\ref{sec:oneDLA}.

\begin{table*}
    \begin{tabular}{c|c|c|c|c}
         &  & $\beta = 0.40$ & $\beta = 0.60$ & $\beta$ free\\
         \hline
         \multirow{7}{1cm}{Coupled DLAs} & $A$ & $2.643\pm 0.003$ & $2.684 \pm 0.003$ & $2.608\pm 0.010$\\
         & $\beta$ & $0.40$ & $0.60$ & $0.23\pm0.05$\\
         & $\log\left(\frac{N_{\rm H_{I,6.3186}}}{\text{cm}^{-2}}\right)$ & $20.565\pm 0.003$ & $20.560\pm 0.002$ & $20.568 \pm 0.003$\\
         & $\log\left(\frac{N_{\rm H_{I,6.3118}}}{\text{cm}^{-2}}\right)$ & $20.945\pm 0.003$ & $20.940 \pm 0.002$ & $20.948 \pm 0.003$\\
         & $x_{\rm H_I}$ & $<0.04$ & $< 0.03$ & $<0.08$\\
         & $\chi^2$ & $5596.87$ & $5640.9$ &$5587.0$\\
         & red. $\chi^2$ & $3.55$ & $3.58$ & $3.55$\\
         & d.o.f. & $1569$ & $1569$ & $1568$\\
         \hline
         \multirow{7}{1cm}{Uncoupled DLAs} & $A$ & $2.634\pm 0.004$ & $2.670\pm 0.004$ & $2.621\pm 0.011$\\
         & $\beta$ & $0.40$ & $0.60$ & $0.32 \pm 0.06$\\
         & $\log\left(\frac{N_{\rm H_{I,6.3186}}}{\text{cm}^{-2}}\right)$ & $20.81 \pm 0.04$ & $20.88\pm 0.04$ & $20.79\pm 0.06$\\
         & $\log\left(\frac{N_{\rm H_{I,6.3118}}}{\text{cm}^{-2}}\right)$ & $20.67^{+0.08}_{-0.11}$ & $20.48^{+0.12}_{-0.18}$ & $20.71^{+0.08}_{-0.12}$\\
         & $x_{\rm H_I}$ & $<0.15$ & $<0.13$ & $<0.28$\\
         & $\chi^2$ & $5583.5$ & $5597.23$ & $5595.6$\\
         & red. $\chi^2$ & $3.54$ & $3.55$ & $3.55$\\
         & d.o.f. & $1568$ & $1568$ & $1567$\\
         \hline    
    \end{tabular}
    \caption{Modeling results for two DLAs and different spectral indices, with uncertainties and upper limits on the parameters corresponding to 1-$\sigma$ and 3-$\sigma$ levels, respectively. The normalization is given in $10^{-17}~\ergscmA$ at 10,000 \AA. The $\chi^2$, reduced $\chi^2$, and degrees of freedom associated with each set of best fit parameters is given as well.}
    \label{tab:results}
\end{table*}

\subsection{Parameter results}
In our analysis, we perform a joint fit of the host ISM column density and neutral fraction parameters.
If the posterior distribution for a parameter is most densely populated around a non-zero value, we estimate the best fit parameter to be the 50\% centile, and use the 16\% and 84\% centiles for the 1-$\sigma$ uncertainties.
If instead the posterior distribution is most densely populated around 0, we report the 3-$\sigma$ upper limit using the 99.7\% centile.

\begin{figure}
    \centering
    \includegraphics[width = \linewidth,trim={0 120 0 140},clip]{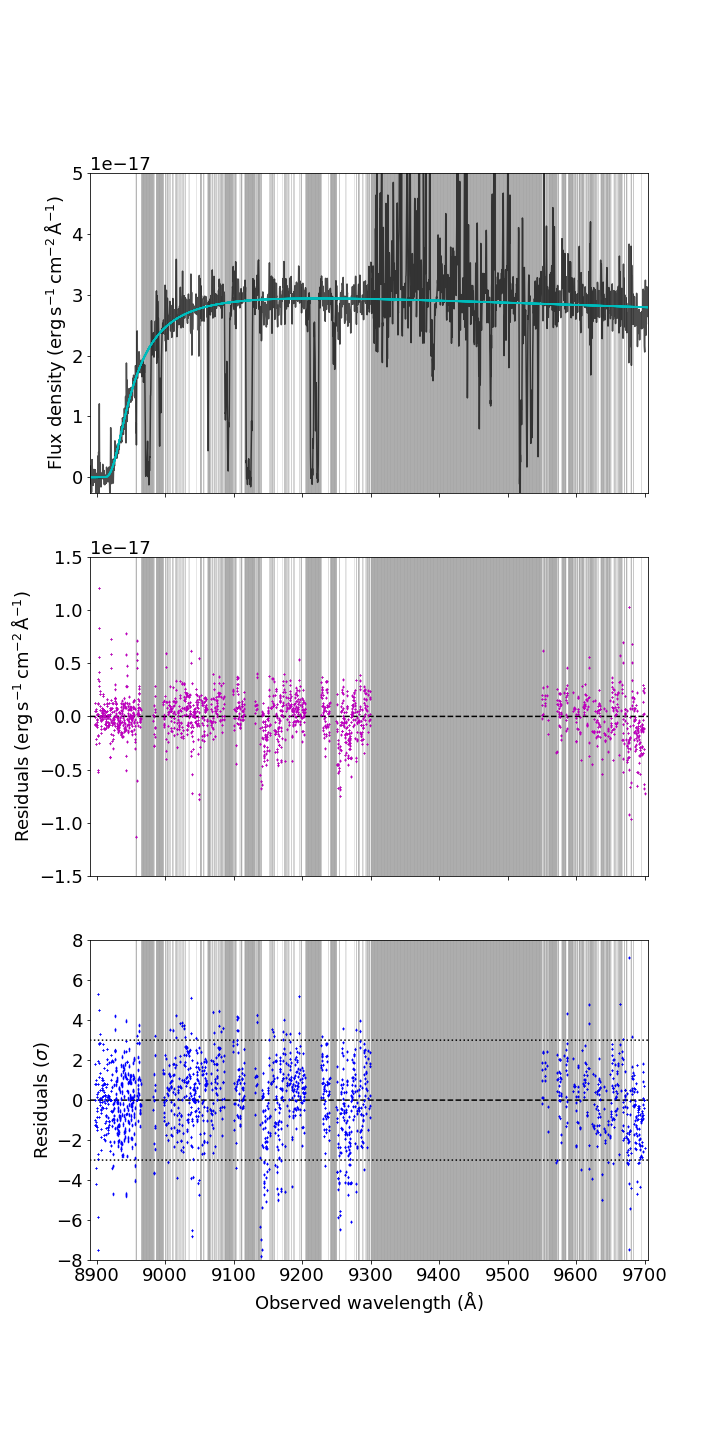}
    \caption{Example fit for the model with coupled DLA column densities. This fit assumes two DLAs at $z=6.3118$ and $z=6.3186$, and a spectral index of $\beta = 0.60$. Regions with metal or telluric lines are shaded in grey and were excluded from analysis. \textbf{Top:} VIS spectral data (black) with the 100 final positions of the walkers (blue). \textbf{Middle:} Residual plot in $\ergscmA$. The dashed black line represents 0 flux. \textbf{Bottom:} Residual plot in $\sigma$; the dashed and dotted black lines represent 0 and 3$\sigma$, respectively.}
    \label{fig:coupled}
\end{figure}

\begin{figure}
    \centering
    \includegraphics[width = \linewidth,trim={15 15 35 15},clip]{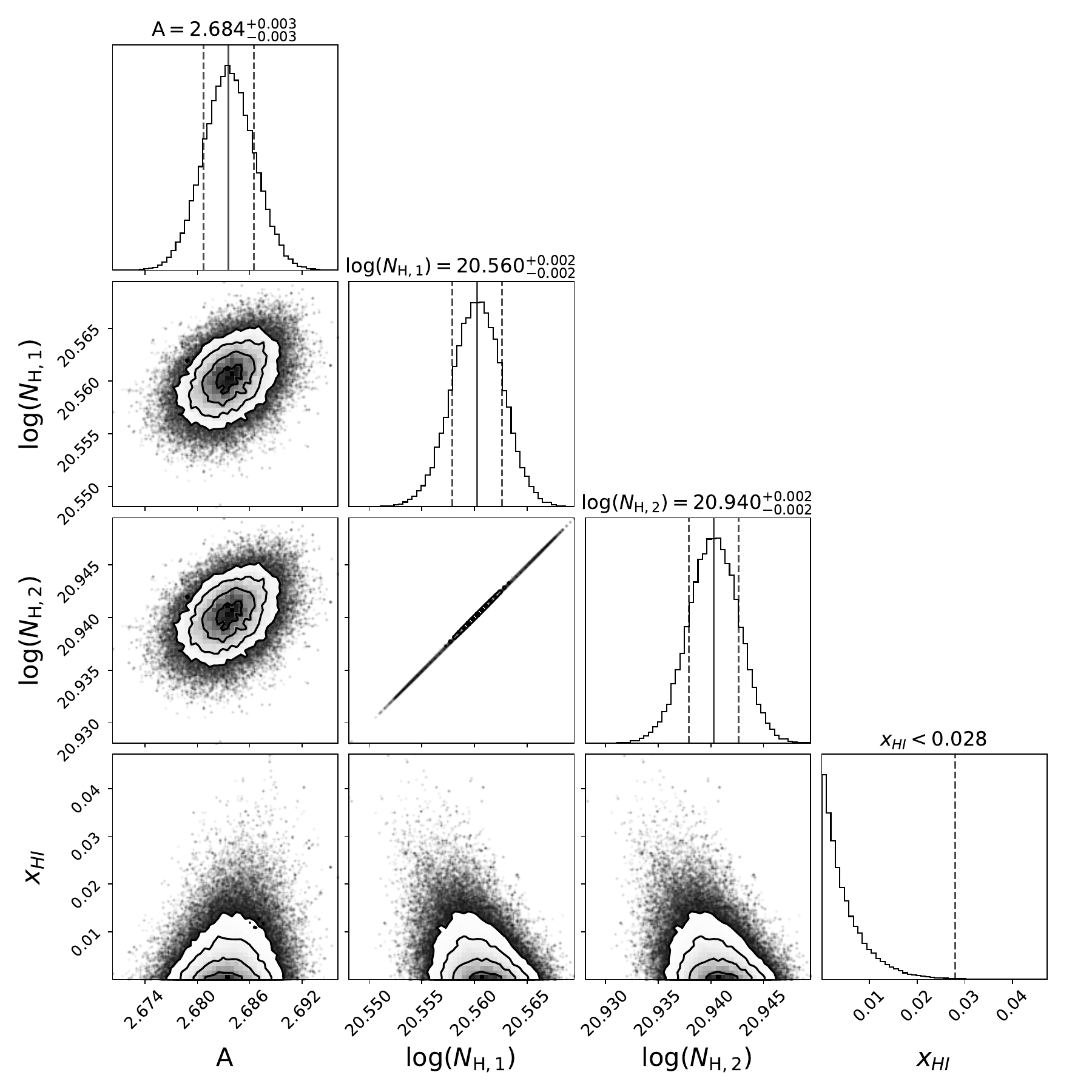}
    \caption{Corner plot for the fit shown in Figure \ref{fig:coupled}. The solid grey lines represent the 50\% centile, while the dashed grey lines show the 1-$\sigma$ uncertainty ranges (16\% and 84\% centiles), or the 3-$\sigma$ upper limit (99.7\% centile). The normalization is given in $10^{-17}\,\ergscmA$. $N_{\rm H,1}$ corresponds to the DLA at $z = 6.3186$, and $N_{\rm H,2}$ corresponds to the DLA at $z = 6.3118$.}
    \label{fig:coupledCorner}
\end{figure}

When fitting the Ly$\alpha$ damping wing of GRB\,210905A with two coupled DLAs, we find that the posterior distribution for $x_{\rm H_I}$ is most densely populated around 0, and obtain upper limits for the neutral fraction for all spectral indices.
When fixing the spectral index to $\beta = 0.60$, as obtained by the SED analysis, we find $\log(N_{\rm H_I, 6.3186}/\text{cm}^{-2}) \sim 20.560 \pm 0.002$, corresponding to $\log(N_{\rm H_I, 6.3118}/\text{cm}^{-2}) \sim 20.940 \pm 0.002$, and a 3-$\sigma$ upper limit of $x_{\rm H_I} < 0.03$ for the neutral fraction.
We find similar results when setting the spectral index to $\beta = 0.40$, with $\log(N_{\rm H_I, 6.3186}/\text{cm}^{-2}) \sim 20.565 \pm 0.003$, corresponding to $\log(N_{\rm H_I, 6.3118}/\text{cm}^{-2}) \sim 20.945\pm 0.003$, and a neutral fraction of $x_{\rm H_I} < 0.04$.
Including the spectral index as a free parameter gives similar results for the two column densities, and an slightly higher upper limit of $x_{\rm H_I} < 0.08$ for the neutral fraction as compared to the fixed spectral index fits.
Figure \ref{fig:coupled} shows the best fit for the coupled DLAs and the spectral index fixed to $\beta = 0.60$, along with the residuals in both flux and $\sigma$ units.
Figure \ref{fig:coupledCorner} shows the posterior distributions and correlations for all parameters of this particular fit, showing that the neutral fraction does not significantly deviate from zero.

When uncoupling the two DLAs, the column densities of the two DLAs are similar, with $\Delta \log(N_{\rm H_I}/\text{cm}^{-2})\sim 0.1-0.4$. For example, for a fixed spectral index of $\beta = 0.6$, the column densities for the DLAs at $z = 6.3186$ and $z = 6.3118$ are $\log(N_{\rm H_I, 6.3186}/\text{cm}^{-2}) \sim 20.88 \pm 0.04$ and $\log(N_{\rm H_I, 6.3118}/\text{cm}^{-2}) \sim 20.48^{+0.12}_{-0.18}$.
While the H$_{\rm I}$ column densities are similar, we consistently obtain a higher column density for the DLA at $z = 6.3186$, while the \citet{Saccardi2023} analysis finds that the metal column density for the DLA at $z = 6.3118$ is higher. However, this does not have a significant effect on the results for the neutral fraction.

\begin{figure}
    \centering
    \includegraphics[width = \linewidth,trim={0 120 0 140},clip]{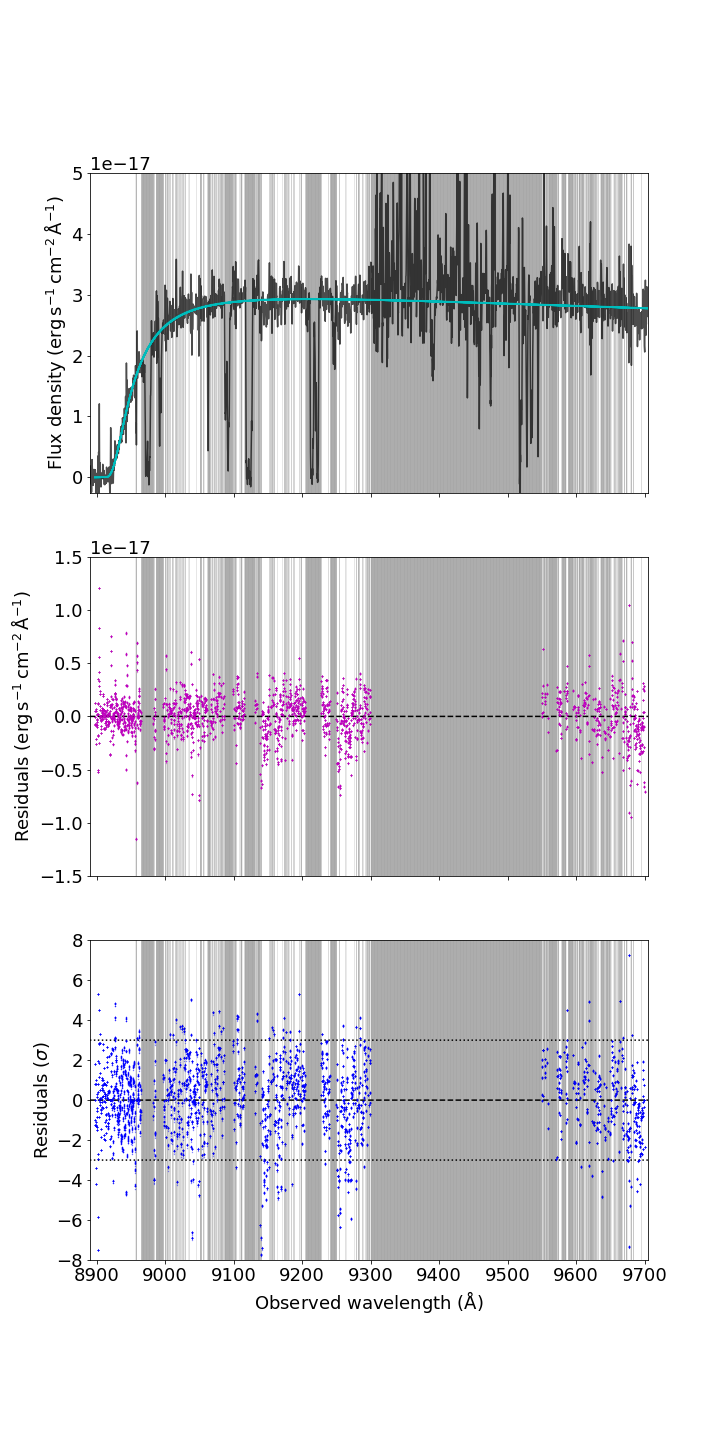}
    \caption{Example fit for a fit with uncoupled column densities for the two DLAs, and a fixed spectral index of $\beta = 0.60$. See Figure \ref{fig:coupled} for description of the three panels.}
    \label{fig:uncoupled}
\end{figure}

\begin{figure}
    \centering
    \includegraphics[width = \linewidth,trim={15 15 30 12},clip]{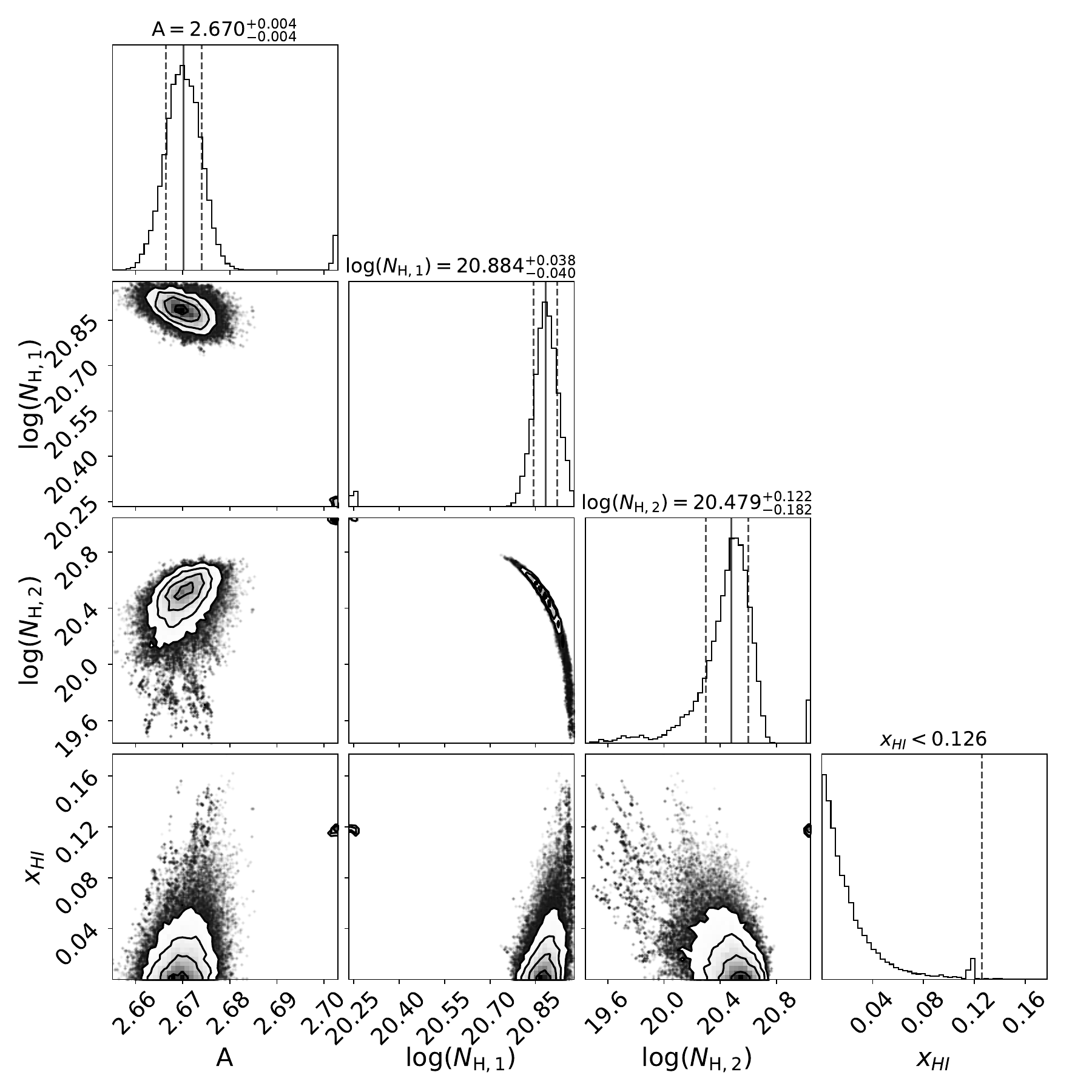}
    \caption{Corner plot for the fit shown in Figure \ref{fig:uncoupled}. The solid grey lines represent the 50\% centile, while the dashed grey lines show the 1-$\sigma$ error ranges (16\% and 84\% centiles). The normalization is given in $10^{-17}\,\ergscmA$. $N_{\rm H,1}$ corresponds to the DLA at $z = 6.3186$, and $N_{\rm H,2}$ corresponds to the DLA at $z = 6.3118$.}
    \label{fig:uncoupledCorner}
\end{figure}

\begin{figure}
    \centering
    \includegraphics[width = \linewidth,trim={0 120 0 140},clip]{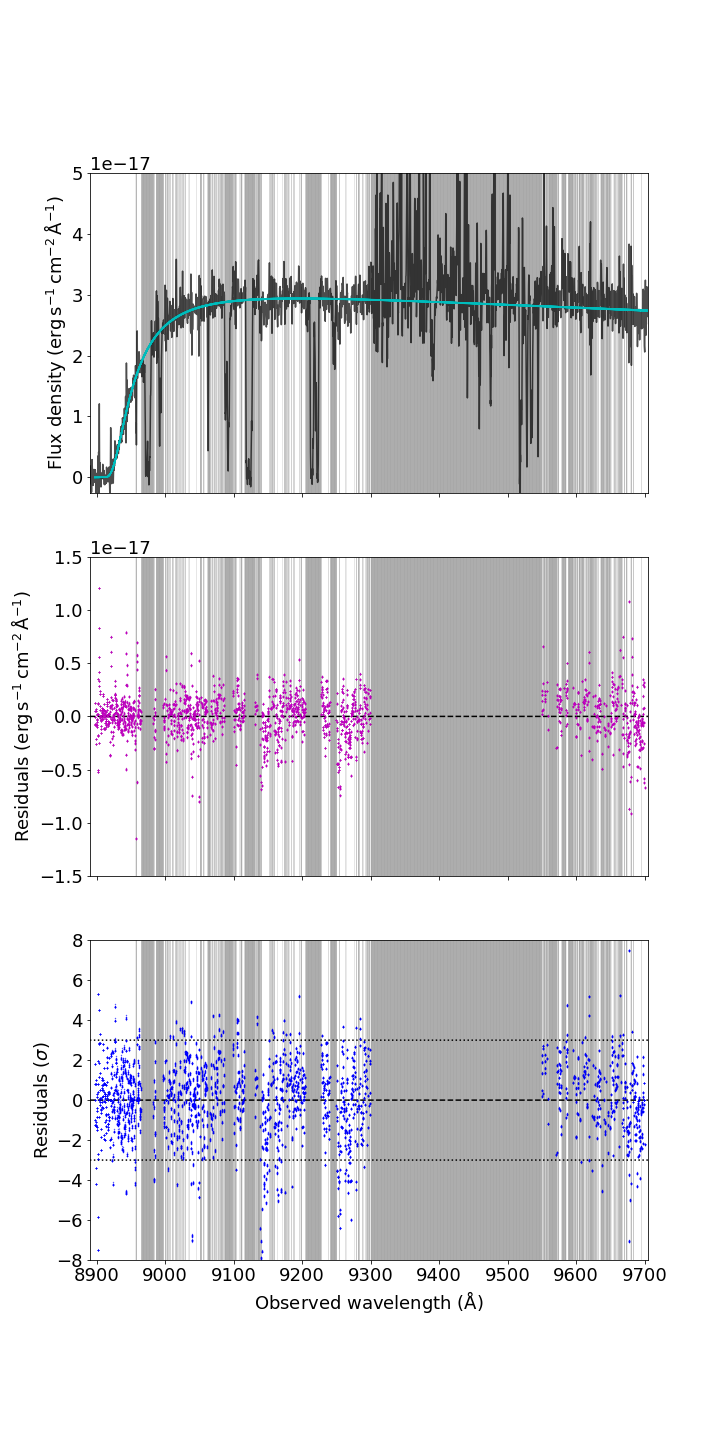}
    \caption{Example fit for a fit with uncoupled column densities for the two DLAs, and a free spectral index with a prior of $0 \leq \beta \leq 2$. See Figure \ref{fig:coupled} for description of the three panels.}
    \label{fig:freebeta}
\end{figure}

\begin{figure}
    \centering
    \includegraphics[width = \linewidth,trim={10 9 20 12},clip]{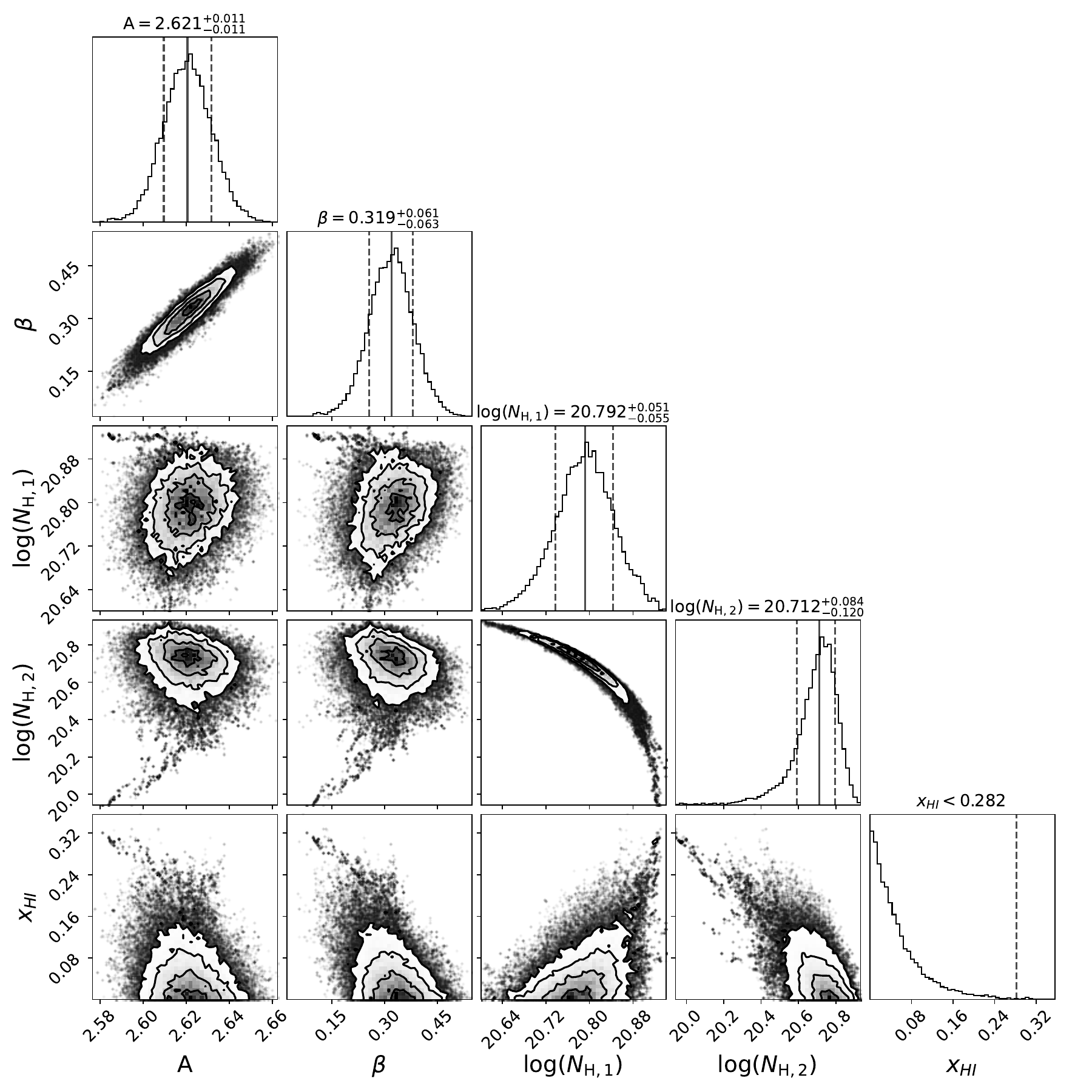}
    \caption{Corner plot for the fit shown in Figure \ref{fig:freebeta}. The solid grey lines represent the 50\% centile, while the dashed grey lines show the 1-$\sigma$ error ranges (16\% and 84\% centiles). The normalization is given in $10^{-17}\,\ergscmA$. $N_{\rm H,1}$ corresponds to the DLA at $z = 6.3186$, and $N_{\rm H,2}$ corresponds to the DLA at $z = 6.3118$.}
    \label{fig:freebeta_corner}
\end{figure}

When uncoupling the DLAs, the posterior distribution for the neutral fraction does still not significantly deviate from 0, but the upper limits increased, ranging from a 3-$\sigma$ upper limit of $x_{\rm H_I} < 0.13$ for a fixed spectral index of $0.60$ (see Figures \ref{fig:uncoupled} and \ref{fig:uncoupledCorner}) up to as high as $x_{\rm H_I} < 0.28$ when including the spectral index as a free parameter (see Figures \ref{fig:freebeta} and \ref{fig:freebeta_corner}).
All corner plots were generated using the \texttt{corner} package \citep{corner}.
All results are summarized in Table~\ref{tab:results}.

\subsection{Parameter correlations}
For both DLA models, the neutral fraction 3-$\sigma$ upper limit tends to decrease as the spectral index increases. 
In the case of the coupled DLAs, the column densities do not seem to depend on spectral index. In the uncoupled DLA case, the column density of the highest-redshift DLA increases with spectral index, while the lowest-redshift DLA shows the opposite trend. However, the sum of the column densities is stable. 
The anti-correlation between the column densities of the two uncoupled DLAs can also be seen in Figure~\ref{fig:coupledCorner}.
This anti-correlation is expected as the two DLAs have the same effect on the shape of the damping wing.

\section{Model Comparison}
\label{sec:compare}
We performed fits of the X-shooter spectrum using models that differ in spectral index and DLA coupling.
As shown in the previous section, we obtain qualitatively similar results for the different model assumptions, but there are quantitative differences for the neutral fraction depending on our model input parameters.
Here we use the python software package \texttt{harmonic} \citep{McEwen2021} to compare the evidence for the different models using the Bayes factor.

The Bayes factor is used for identifying a preferred model by comparing posterior probabilities \citep{Jeffreys1935, Jeffreys1939, Kass1995}, and is essentially a comparison of evidence for two models. The python software package \texttt{harmonic} is an implementation of the learnt harmonic mean estimator \citep{Newton1994, McEwen2021}. \citet{Newton1994} found that the harmonic mean of the likelihoods of posterior chains could be used to estimate the marginal likelihood, or evidence, for a model. While the original harmonic mean estimator could be unreliable, \texttt{harmonic} uses the alternative \citet{Gelfand1994} harmonic mean estimator, which improved upon the original one by introducing a target distribution \citep{McEwen2021}. \texttt{harmonic} uses the posteriors of a fit to estimate the optimal target distribution using machine learning, and to compute the Bayesian evidence and posterior probability for a model \citep{McEwen2021}. 

\subsection{Model Selection and Preferred Model Results}
The marginal likelihood is an average likelihood for a model based off of the probability of generating the data given the prior \citep{McEwen2021}.
Using \texttt{harmonic}, we find the estimated natural-log of the marginal likelihood, $\ln(ML)$, for each model using the posterior chains. The natural-log of the Bayes factor comparing model 1 to model 0 is defined as $\ln(B_\mathrm{10}) = \ln(ML_\mathrm{1}) - \ln(ML_\mathrm{0})$.
The estimated marginal likelihoods for each of the two DLA models can be found in Table \ref{tab:bayes}.

\begin{table}
    \centering
    \begin{tabular}{c|c|c|c}
          & $\beta = 0.4$ & $\beta = 0.6$ & $\beta$ free \\
          \hline
         Coupled DLAs & $-2809.8$ & $-2832.6$ & $-2804.9$\\ 
         Uncoupled DLAs & 
         $-2797.7$ & $-2804.0$ & $-2795.3$
    \end{tabular}
    \caption{The estimated marginal likelihoods of each 2-DLA model using the \texttt{harmonic} python software package.}
    \label{tab:bayes}
\end{table}

\begin{figure}
    \centering
    \includegraphics[width = \linewidth,trim={0 120 0 140},clip]{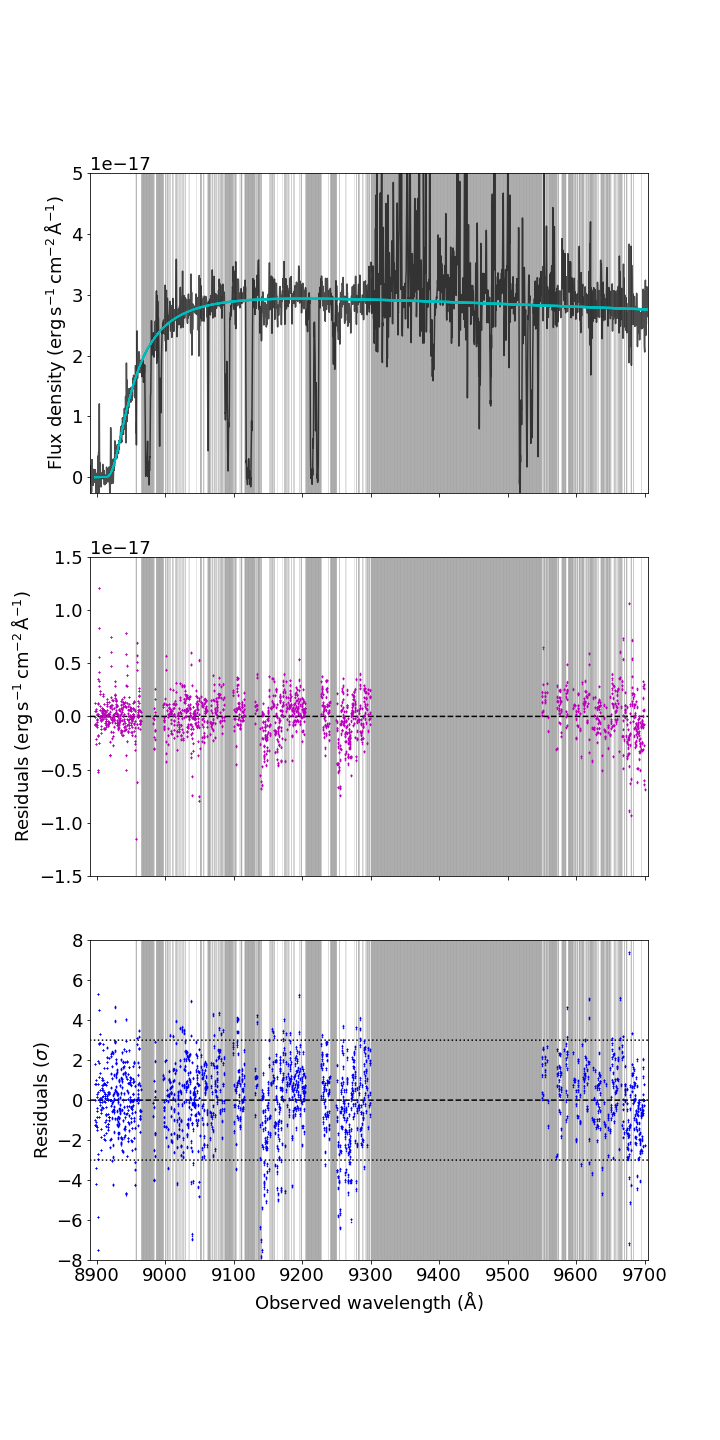}
    \caption{Fit for the statistically preferred model with uncoupled column densities for the two DLAs and a fixed spectral index of $\beta = 0.40$. See Figure \ref{fig:coupled} for description of the three panels.}
    \label{fig:bestfitplot}
\end{figure}

\begin{figure}
    \centering
    \includegraphics[width = \linewidth,trim={15 15 30 12},clip]{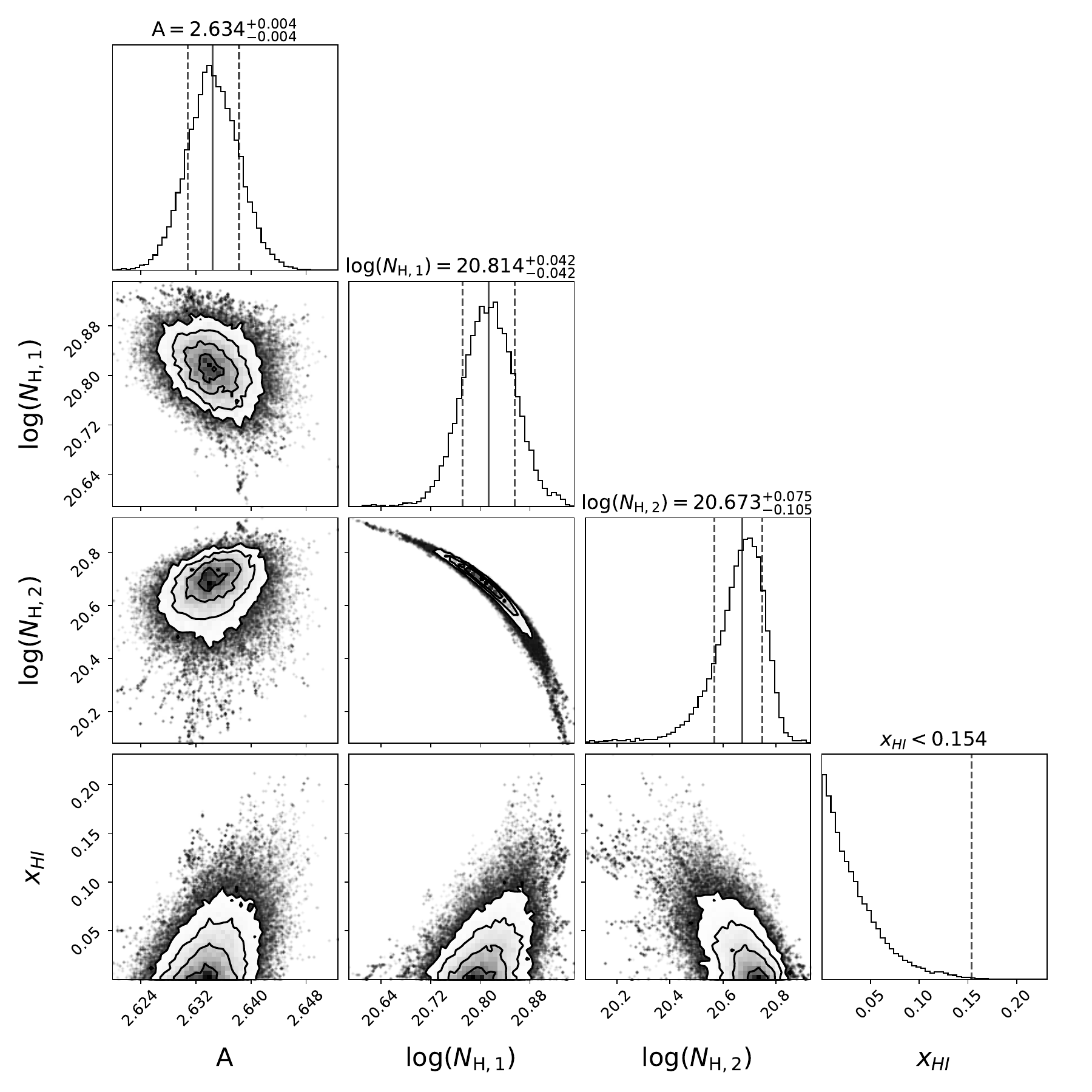}
    \caption{Corner plot for the fit shown in Figure \ref{fig:bestfitplot}. The solid grey lines represent the 50\% centile, while the dashed grey lines show the 1-$\sigma$ error ranges (16\% and 84\% centiles). The normalization is given in $10^{-17}\,\ergscmA$. $N_{\rm H,1}$ corresponds to the DLA at $z = 6.3186$, and $N_{\rm H,2}$ corresponds to the DLA at $z = 6.3118$.}
    \label{fig:best_fit_corner}
\end{figure}

Based on the estimated Bayes factors, we find that the model with the coupled DLA column densities is generally not preferred. When taking the model with coupled column densities to be model 0 and the model with uncoupled column densities to be model 1, $\ln(B_\mathrm{10})$ ranges from $\sim 11.6$ to $\sim 28.6$. According to \citet{Kass1995},  $2\ln(B_\mathrm{10}) > 10$ is very strong evidence against model 0, so the model with coupled column densities model is strongly disfavored.
A lower spectral index is also generally preferred.
When comparing the different uncoupled DLA fits, we find $\ln(B_\mathrm{10}) \sim 6.3$ when comparing the models with their spectral indices fixed to $0.60$ and $0.40$, and $\ln(B_\mathrm{10}) \sim 2.4$ when comparing the model with the spectral index fixed to $0.40$ to the model with a free spectral index, which found a spectral index of $\beta = 0.32\pm0.06$. The fit with the spectral index fixed to $0.60$ is strongly not preferred, but the models with a fixed spectral index of $\beta = 0.40$ and a free spectral index are harder to distinguish as $2 < 2\ln(B_\mathrm{10}) < 6$ is considered positive, but not strong, evidence against model 0 \citep{Kass1995}.
Since we are not confident in our ability to constrain the spectral index given the short lever arm, and since there is not strong evidence that the free spectral index fit is preferred over a simpler version of the model where the spectral index is fixed to $0.40$, we take the model with $\beta = 0.40$ and uncoupled DLAs to be the preferred model.
For this model, the neutral fraction does not significantly deviate from zero and has 3-$\sigma$ upper limits of $x_{\rm H_I} \lesssim 0.15$. The fit and posteriors for this fit are shown in Figures \ref{fig:bestfitplot} and \ref{fig:best_fit_corner}, respectively.

\section{Comparison to Ionized Bubble Model Results}
\label{sec:McQuinn}
The \citet{McQuinn2008} model is an approximate version of the \citet{MiraldaEscude1998} model, but it differs from the \citet{MiraldaEscude1998} and \citet{Totani2006} models in the sense that it allows for an ionized bubble of radius $R_{\rm b}$ around the host galaxy, rather than relying on the assumed values for the $z_{\rm IGM,u}$ and $z_{\rm IGM,l}$. We performed fits with the \citet{McQuinn2008} model to check the robustness of our result.

\begin{figure}
    \centering
    \includegraphics[width = \linewidth,trim={0 110 0 150},clip]{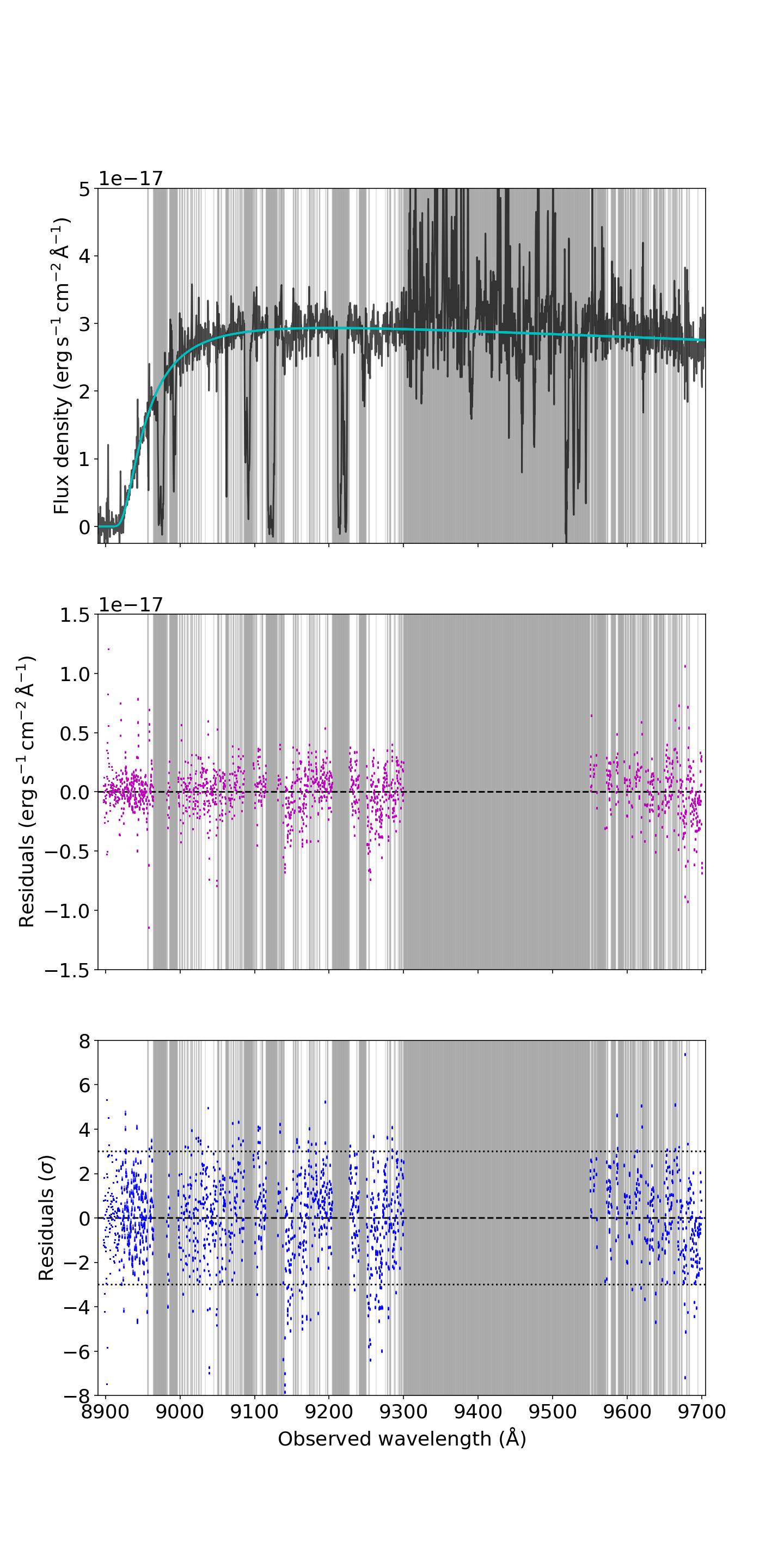}
    \caption{Fit using the \citet{McQuinn2008} model with uncoupled column densities for the two DLAs, a fixed spectral index of $\beta = 0.40$, and an $R_{\rm b}$ upper limit of $60~\text{Mpc}\;h^{-1}$ or $\sim 90~\text{Mpc}$. See Figure \ref{fig:coupled} for description of the three panels.}
    \label{fig:mcquinn60plot}
\end{figure}

\begin{figure}
    \centering
    \includegraphics[width = \linewidth,trim={10 10 28 10},clip]{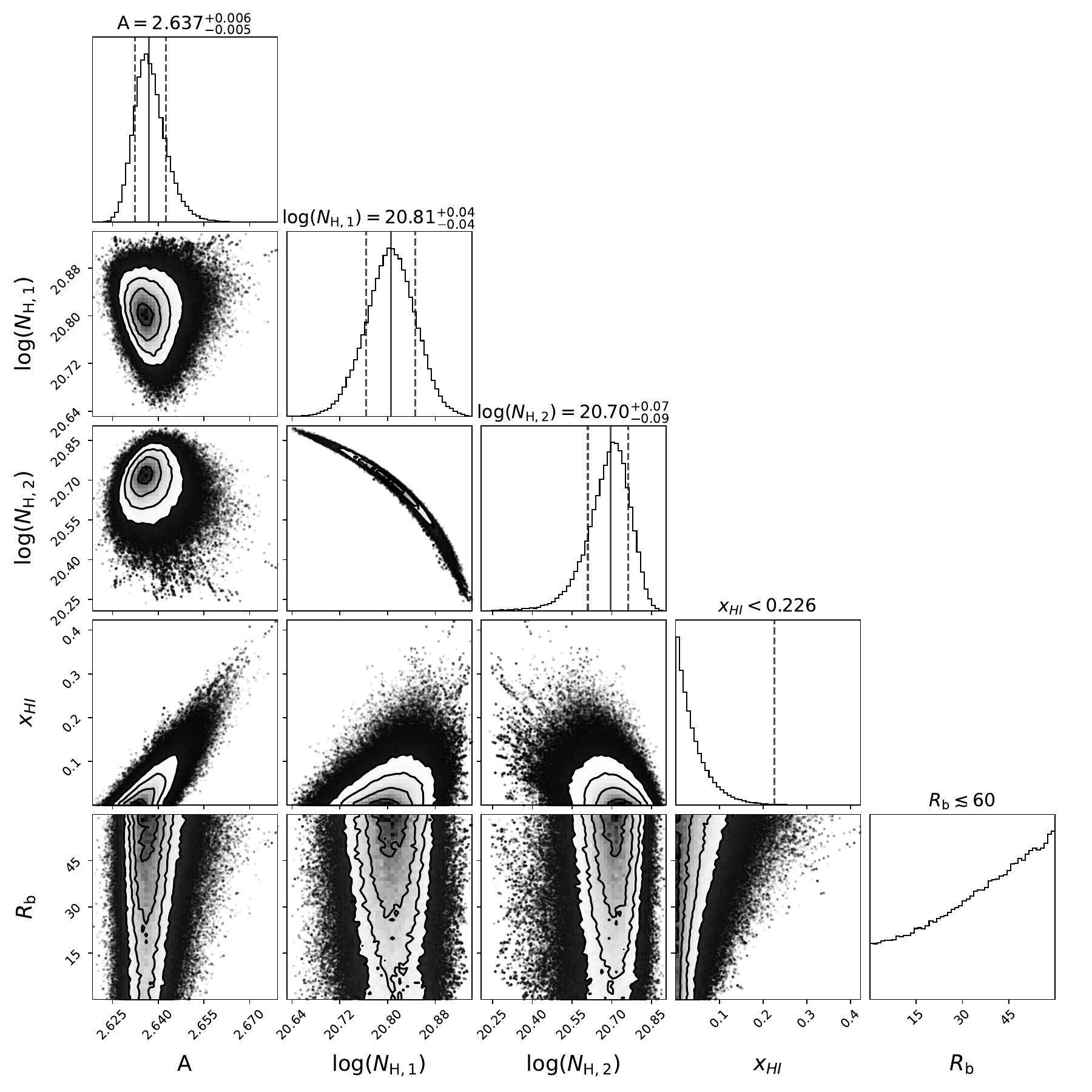}
    \caption{Corner plot for the fit shown in Figure \ref{fig:mcquinn60plot}. The solid grey lines represent the 50\% centile, while the dashed grey lines show the 1-$\sigma$ error ranges (16\% and 84\% centiles). The normalization is given in $10^{-17}\,\ergscmA$ and the ionized bubble radius is given in $\text{Mpc}\;h^{-1}$. $N_{\rm H,1}$ corresponds to the DLA at $z = 6.3186$, and $N_{\rm H,2}$ corresponds to the DLA at $z = 6.3118$.}
    \label{fig:mcquinn60corner}
\end{figure}

In all cases, the \citet{McQuinn2008} model also results in a neutral fraction posterior that does not deviate significantly from 0. 
An example fit and posterior distribution using the statistically preferred setup (uncoupled DLAs, a fixed spectral index of $\beta = 0.40$) for the \citet{McQuinn2008} model give a 3-$\sigma$ upper limit of $x_{\rm H_I} \leq 0.23$ (see Figures \ref{fig:mcquinn60plot} and \ref{fig:mcquinn60corner}). 
This is slightly higher than the result obtained with the \citet{MiraldaEscude1998} model with the same setup (3-$\sigma$ upper limit of $x_{\rm H_I} \leq 0.15$), but consistent.
We note that the DLA column density results are also very similar to those found for uncoupled DLAs when using the \citet{MiraldaEscude1998} model. A comparison of the neutral fraction results for the \citet{MiraldaEscude1998} and \citet{McQuinn2008} models with uncoupled DLAs can be seen in Table \ref{tab:compare}.

\begin{table}
    \centering
    \begin{tabular}{c|c|c|c}
         & $\beta = 0.4$ & $\beta = 0.6$ & $\beta$ free \\
         \hline
         \citet{MiraldaEscude1998} & $<0.15$ & $<0.13$ & $<0.28$\\
         \citet{McQuinn2008} & $<0.23$ & $<0.12$ & $< 0.57$
    \end{tabular}
    \caption{Table comparing results from the \citet{MiraldaEscude1998} and \citet{McQuinn2008} models. The upper limit when fitting for $\beta$ is much larger for the \citet{McQuinn2008} model, but for a fixed spectral index the upper limits from the two models are consistent.}
    \label{tab:compare}
\end{table}

The radius of the comoving ionized bubble is not well constrained, but trends towards the largest allowed radius of $\sim 60~\text{Mpc}\;h^{-1}$ or $\sim 90~\text{Mpc}$. The behavior of this parameter could indicate that there is no clear boundary between the ionized bubble around the host galaxy and the IGM, which would indicate that the IGM is highly ionized along the line of sight.

When performing these fits for larger $R_{\rm b}$ upper bounds the posteriors display the same behavior, with an unconstrained $R_{\rm b}$ tending towards the largest allowed radius. This indicates that the behavior of the $R_{\rm b}$ parameter is not caused by our chosen $R_{\rm b}$ range. When increasing $R_{\rm b}$, the neutral fraction still does not significantly deviate from 0. The upper limits on $x_{\rm H_I}$ do increase with $R_{\rm b}$, but this is likely because the farther away neutral hydrogen is from the redshift associated with the red Ly$\alpha$ damping wing, the less noticeable its impact on the shape of the damping wing and the less precisely it can be constrained. This same effect can be seen when using a shell implementation of the \citet{MiraldaEscude1998} model (see Section \ref{sec:shells}). Fits using $0~\text{Mpc}\;h^{-1} \leq R_{\rm b} \leq 130~\text{Mpc}\;h^{-1}$ and $0~\text{Mpc}\;h^{-1} \leq R_{\rm b} \leq 355~\text{Mpc}\;h^{-1}$ priors can be found in Appendix \ref{sec:additionalMcquinn}.

For completeness, we also performed a fit with the \citet{MiraldaEscude1998} model with $z_{\rm IGM,u}$ fixed to a lower redshift to create an ionized region around the host galaxy similar to the \citet{McQuinn2008} model. 
We use $z_{\rm IGM,u} = 6.23$ corresponding to a bubble radius of $\sim 32~\text{Mpc}\;h^{-1}$. 
This is the bubble radius that is expected for a neutral fraction of $\sim 0.25$, which is the estimated value at $z\sim 6.3$ based on recent EoR models \citep[e.g.,][]{Ishigaki2018, Naidu2020, Lidz2021}.
For uncoupled DLAs and a fixed spectral index of $\beta = 0.4$, we find a 3-$\sigma$ upper limit of $x_{\rm H_I} \lesssim 0.26$, which is consistent with the result obtained with the \citet{McQuinn2008} model.
Given that this result requires a strong assumption about the size of the ionized bubble around the host galaxy, we adopt the 3-$\sigma$ upper limit of $x_{\rm H_I} \lesssim 0.23$ obtained with the \citet{McQuinn2008} model as the preferred result.

\section{Discussion}

In the previous sections we have shown that the preferred models point to a negligible neutral fraction of hydrogen with an upper limit of $x_{\rm H_I} \lesssim 0.15$ and $x_{\rm H_I} \lesssim 0.23$ when using the \citet{MiraldaEscude1998} and \citet{McQuinn2008} models, respectively.
Here we compare our results with those using a shell implementation of the  \citet{MiraldaEscude1998} model, explore how the result would be different if we did not obtain a high-quality spectrum (i.e., we would have only one DLA), and discuss the effects of potential spectral curvature and absorption lines on the neutral fraction estimate.
We also explore the implications of these results in the broader context of the EoR, and for future high-redshift GRB analyses.

\subsection{Effect of $z_{\rm IGM_l}$ for the \citet{MiraldaEscude1998} model}
\label{sec:zlchoice}

The model parameter $z_{\rm IGM_l}$ represents the redshift at which the fraction of neutral hydrogen in the IGM becomes negligible. 
For redshifts between $z_{\rm IGM_l}$ and $z_{\rm IGM_u}$, the fraction of neutral hydrogen is assumed to be constant. 
The choice of $z_{\rm IGM_l}$ can have an impact on the modeling results, as a $z_{\rm IGM_l}$ value closer to the GRB redshift gives a higher neutral fraction measurement. 
However, beyond a certain redshift, $z_{\rm IGM_l}$ becomes sufficiently large that the neutral fraction result is no longer impacted. 
This is likely because beyond a certain redshift, the presence of neutral hydrogen no longer has a discernible effect on the damping wing.

\begin{figure}
    \centering
    \includegraphics[width = \linewidth]{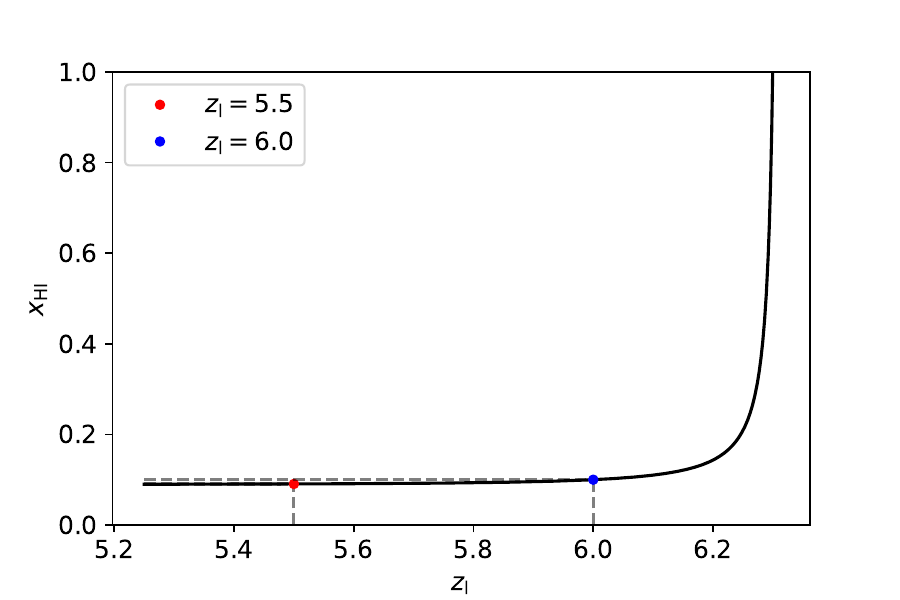}
    \caption{Figure showing impact of $z_{\rm IGM_l}$ on neutral fraction. In this case, $z_{\rm IGM_l}$ does not have a large impact on the neutral fraction result as long as $z_{\rm IGM_l} \lesssim 6.2$.}
    \label{fig:zleffect}
\end{figure}

To test the impact of our choice of $z_{\rm IGM_l}$ we simulated a Ly$\alpha$ damping wing using $z_{\rm IGM_l} = 6.0$, $z_{\rm DLA} = 6.3118$, $z_{\rm host} = 6.3186$, $\beta = 0.5$, $\log(N_{\rm H_I}/\text{cm}^{-2}) = 21$, and $x_{\rm H_I} = 0.10$. 
We fit the simulated data with fixed $z_{\rm IGM_l}$ values ranging from $z=5.25$ to $z=6.31$, to assess how the result for the neutral fraction changes with $z_{\rm IGM_l}$; see Figure \ref{fig:zleffect}. 
It is evident that our choice of $z_{\rm IGM_l}$ does not impact the retrieved neutral fraction value, and choosing $z_{\rm IGM_l} = 5.5$ to correspond to a late ending reionization would still result in a very similar neutral fraction value. Therefore, our initial choice of $z_{\rm IGM_l}$ did not impact the analysis presented in this paper, and neutral fraction results using the \citet{MiraldaEscude1998} method are independent of the chosen reionization model as long as $z_{\rm IGM_l}$ is sufficiently far from the GRB redshift ($z_{\rm IGM_l} \lesssim 6.2$).

\subsection{\citet{MiraldaEscude1998} model with shells}
\label{sec:shells}
To better account for the patchiness of reionization, we implemented the \citet{MiraldaEscude1998} model in a series of shells around the GRB host galaxy. The shell implementation has independent contributions to the shape of the damping wing from shells with widths of $\Delta z = 0.1$ from the GRB redshift out to $z = 5.5$. This allows more flexibility to explore any evolution in the ionization state of the IGM, but may also be pushing the limits of what can be done with the data.

We perform a fit using the statistically preferred assumptions (uncoupled DLAs and a fixed spectral index of $\beta = 0.4$), but with a 10000 step burn-in and 20000 step production chain. 
We find that the neutral fraction always tends towards 0, but the posterior gradually becomes flatter for shells further away from the GRB redshift. 
This is expected as neutral hydrogen farther away from the GRB has a less pronounced effect on the Ly$\alpha$ damping wing and becomes harder to constrain. 
For shells $z = 6.3186-6.2$, $z = 6.2-6.1$, and $z = 6.1-6.0$ we find 3-$\sigma$ upper limits of $x_{\rm H_I} \lesssim 0.23$, $x_{\rm H_I} \lesssim 0.39$, and $x_{\rm H_I} \lesssim 0.79$, respectively. 
For $z = 6.0 - 5.7$, the neutral fraction is most densely populated around 0, but no meaningful upper limit can be given. 
For lower redshifts the posterior still tends slightly towards 0, but is mostly flat. 
The fit and corresponding corner plot can be seen in Figures \ref{fig:shellFit} and \ref{fig:shellCorner}, respectively. 
These results point toward a largely ionized medium around GRB210905A, as was also found using the original \citet{MiraldaEscude1998} model and the \citet{McQuinn2008} model.

\subsection{Modeling with One DLA}
\label{sec:oneDLA}
\begin{figure}
    \centering
    \includegraphics[width = \linewidth,trim={0 120 0 120},clip]{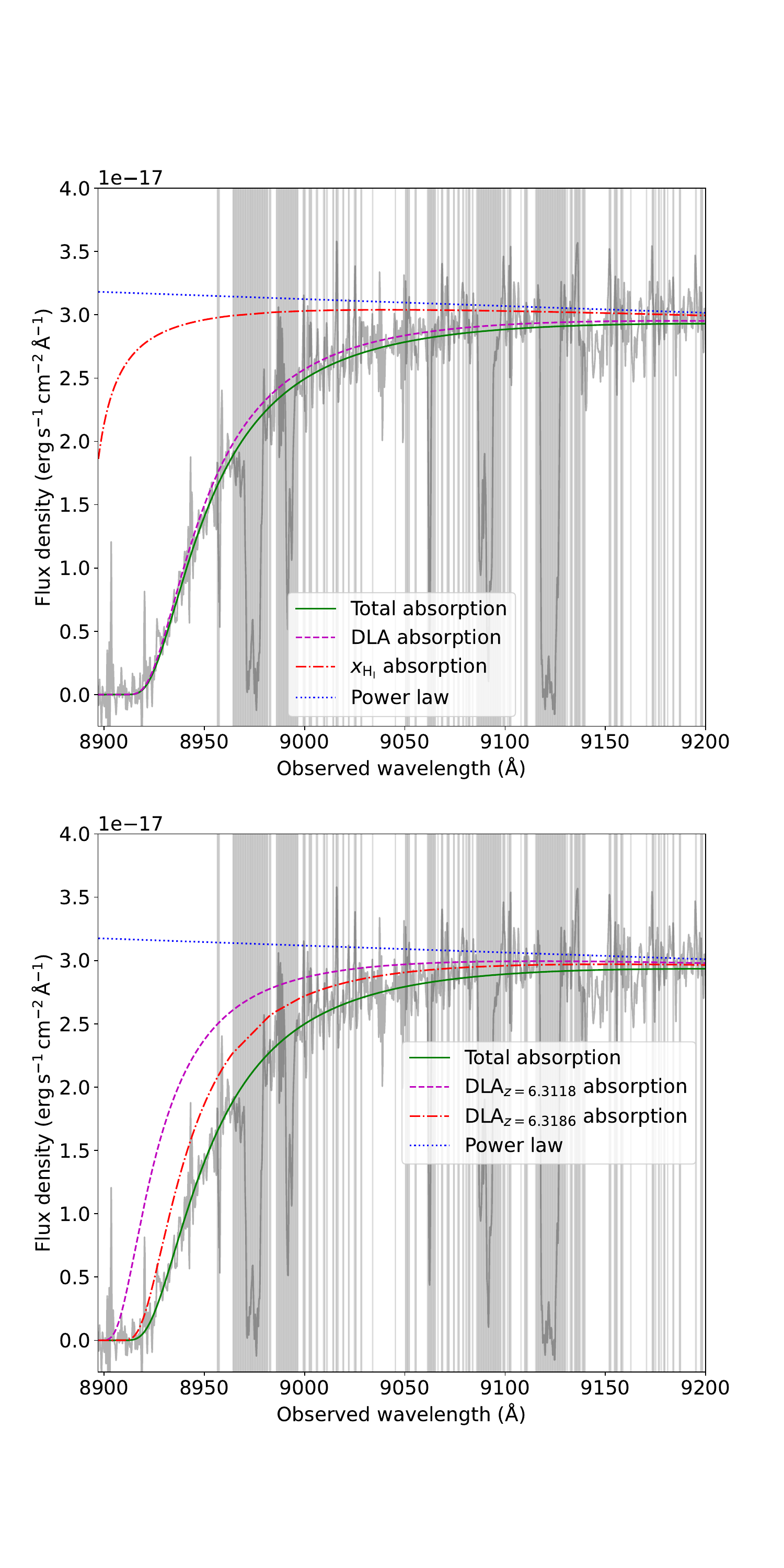}
    \caption{Fit comparing the contributions of different absorbers for the one DLA and two uncoupled DLA cases. \textbf{Top:} Fit for one DLA at $z=6.3186$ and a fixed spectral index of $\beta = 0.40$ with the original power law (dotted blue), the individual contributions from the IGM (dash-dotted red) and the DLA (dashed purple), and the total absorption (solid green). \textbf{Bottom:} Fit for two uncoupled DLAs and a fixed spectral index of $\beta = 0.40$ with the original power law (dotted blue), the individual contributions from the DLA at $z = 6.3118$ (dash-dotted red) and the DLA at $z = 6.3186$ (dashed purple), and the total absorption (solid green).}
    \label{fig:oneDLAplot}
\end{figure}

\begin{figure}
    \centering
    \includegraphics[width = \linewidth,trim={15 15 25 10},clip]{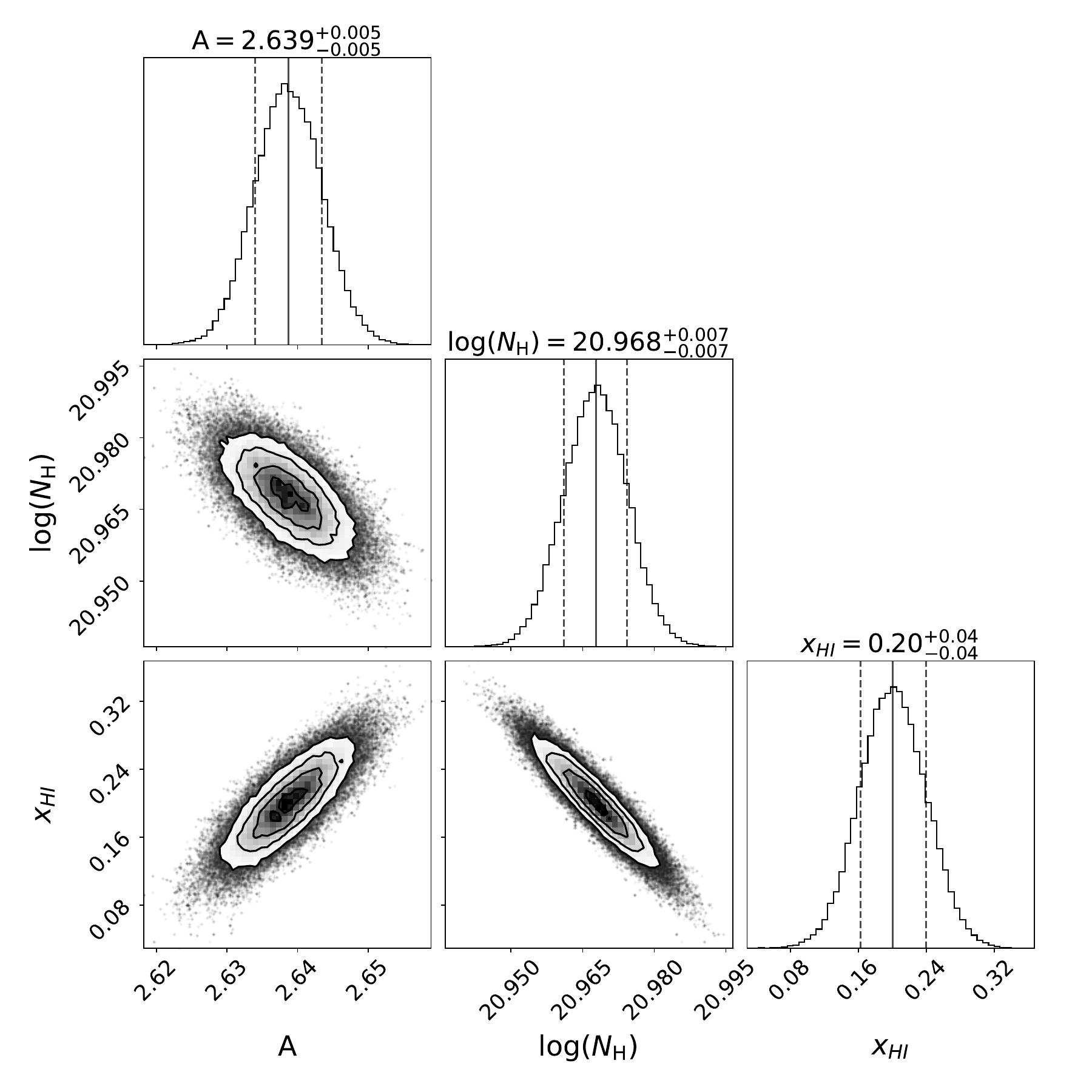}
    \caption{Corner plot for the fit shown in Figure \ref{fig:oneDLAplot}. The solid grey lines represent the 50\% centile, while the dashed grey lines show the 1-$\sigma$ error ranges (16\% and 84\% centiles). The normalization is given in $10^{-17}\,\ergscmA$. $N_{\rm H,1}$ corresponds to the DLA at $z = 6.3186$.}
    \label{fig:oneDLAcorner}
\end{figure}

The modeling with two DLAs shows that the neutral fraction does not significantly deviate from zero.
However, if the spectrum had a significantly lower spectral resolution and/or signal to noise, the absorption lines from the two DLAs would have been indistinguishable, and we would have fit the damping wing assuming one DLA.
When fitting for only one DLA, the neutral fraction deviates from $0$ for all spectral indices, with results ranging from $x_{\rm H_{I}} = 0.08\pm0.04$ for a fixed spectral index of $0.60$ to $x_{\rm H_{I}} = 0.29 \pm 0.06$ when the spectral index is treated as a free parameter; see Table \ref{tab:oneDLA} for the full results.
However, it makes sense that the neutral fraction increases when one of the DLAs is removed, since there is now one less component contributing to the damping wing, so the neutral fraction must increase to create the same shape through the wing (see Figure~\ref{fig:oneDLAplot}).  The posterior distribution for one DLA at $z = 6.3186$ and a spectral index fixed to $\beta = 0.40$ is shown in Figure \ref{fig:oneDLAcorner}.

\begin{table}
    \centering
    \begin{tabular}{c|c|c|c}
        & $\beta = 0.40$ & $\beta = 0.60$ & $\beta$ free\\
        \hline
        $A$ & $2.639 \pm 0.005$ & $2.668 \pm 0.005$ & $2.619\pm 0.012$\\
        $\beta$ & $0.40$ & $0.60$ & $0.26 \pm 0.08$\\
        $\log\left(\frac{N_{\rm H_{I,6.3186}}}{\text{cm}^{-2}}\right)$ & $20.968 \pm 0.007$ & $20.982\pm 0.006$ & $20.957 \pm 0.009$\\
        $x_{\rm H_I}$ & $0.20\pm 0.04$ & $0.08\pm 0.04$ & $0.29\pm 0.06$\\
        $\chi^2$ & $5594.1$ & $5609.6$ & $5590.9$\\
        red. $\chi^2$ & $3.55$ & $3.56$ & $3.55$\\
        d.o.f & $1569$ & $1569$ & $1568$\\
         \hline 
    \end{tabular}
    \caption{Modeling results for different spectral indices assuming one DLA at $z = 6.3186$, with parameter uncertainties given at 1-$\sigma$ levels. The normalization is given in $10^{-17}~\ergscmA$ at 10,000 \AA. The $\chi^2$, reduced $\chi^2$, and degrees of freedom associated with each set of best fit parameters is given as well.}
    \label{tab:oneDLA}
\end{table}

This highlights the importance of using instruments capable of obtaining a decent spectral resolution at wavelengths covering the Ly$\alpha$ damping wing for GRBs at $z \sim 6-10$, such as X-shooter, for GRB follow-up observations. It is clear in the analysis by \citet{Saccardi2023} that there are two velocity systems belonging to the same GRB host galaxy complex, and both are close enough to impact the shape of the Ly$\alpha$ damping wing. Had we not been aware of the second DLA, we would have derived an inflated neutral fraction in the IGM around the GRB host galaxy. Future instruments such as SCORPIO \citep{SCORPIO}, which includes a long-slit spectrograph, will be even better for GRB follow-up. Since long-slit spectrographs are easier to flux-calibrate, it could eliminate some of the issues that may arise when fitting for $\beta$.

\subsection{Spectral Indices and Curvature}
\label{sec:spectralIndexCorrelation}
In our analysis, we fit the data with multiple fixed spectral index values, because the lack of a sufficient lever arm makes it difficult to fit for the spectral index.
In Table \ref{tab:results}, we show that in general both the fit quality and the neutral fraction increase with a lower spectral index. 
Based on the latter, one may prefer the $\beta \sim 0.4$ fits, which is based on extrapolation from the X-ray spectrum with an assumed cooling break. 
However, \citet{Rossi2022} find an optical spectral index of $\beta_{\rm opt} = 0.60 \pm 0.04$, and argue that the cooling break occurs below but close to the X-ray energy range. 
We would like to note that the cooling break does not result in a sharp transition between two spectral indices, but instead is a more gradual change potentially over multiple orders of magnitude in wavelength \citep{Granot2002,vanEerten2009,Uhm2014}. Therefore, even if the cooling break is relatively far away in wavelength from either the optical or X-ray regime, it may still have an effect on the derived spectral indices, in which case the neutral fraction estimate would be less certain.

\subsection{Impact of Telluric Spectral Lines}
\label{sec:absorption}
For GRB\,210905A, there is a large number of absorption lines throughout the damping wing, particularly in the upper part of the wing where a higher neutral fraction has the biggest impact.
This makes it more difficult to constrain the neutral fraction confidently.
For GRBs with a similar redshift, a significant portion of the upper  damping wing will be contaminated by telluric emission and absorption lines present around $\sim 9,000$~\AA.
Having a larger sample of high-redshift GRBs at a range of different redshifts will alleviate this issue of telluric spectral lines in a crucial part of the damping wing. 
This is obviously not an issue for space-based observatories such as JWST \citep{Gardner2006}.

\subsection{Broader context}

The neutral fraction around GRB\,210905A is generally in agreement with other neutral fraction analyses \citep[e.g.,][]{Totani2006, Fan2006, Ouchi2010, Chornock2013, Hartoog2015, Ouchi2018}.
However, some more recent results are in moderate disagreement. 
For example, \citet{Zhu2024} stack Ly$\alpha$ transmission profiles based on dark gaps in Ly$\beta$ transmission to find evidence of damping wing-like features.
They find a neutral fraction of $x_{\rm H_I} \geq 0.061 \pm 0.039$ at $z\sim 5.8$, suggesting that some dark gaps at this redshift are caused by neutral islands.
A similar analysis by \citet{Spina2024} finds $x_{\rm H_I} < 0.44$ at $z \sim 5.9$.
The estimate for the fraction of neutral hydrogen around the host galaxy of GRB\,210905A was complicated by a range of possible spectral indices and absorption lines in a key portion of the Ly$\alpha$ damping wing.
There were also multiple DLAs near the GRB redshift, which can make it more difficult to disentangle the contributions of each component \citep{Heintz2024}.
Measuring the neutral fraction around the end of the EoR is also difficult.
\citet{Lidz2021} discuss how estimating the neutral fraction is easier at higher redshifts where the ionized bubble around the host galaxy is smaller and changes in $x_{\rm H_I}$ have a greater impact on the shape of the damping wing.
Finally, it is always possible that the host galaxy of GRB\,210905A falls in a region of space with an unusually large inoized region along the line of sight, and does not fully represent the global neutral fraction at this redshift. 
This is commonly a concern for quasars since they are usually found in massive dark matter halos, which tend to fall in regions of space that are already highly ionized \citep{Alvarez2007,Lidz2007, Davies2018}.
The result for the neutral fraction around GRB~210905A is one data point among many that can help improve our understanding of the Epoch of Reionization, but it cannot in itself be used to support or rule out any Reionization models. Increasing the sample of high-redshift GRBs and other high-redshift probes will be vital to increasing our understanding of the EoR (see Section \ref{sec:highzgrbs}).

\subsection{Sightline Analysis at $z\sim 6.3$}

To determine the likelihood of encountering an over-ionized sightline for GRB\,210905A, we perform a basic simulation to determine the number of fully ionized sightlines for an IGM with $x_{\rm H_I} \sim 0.25$, as would be expected for a redshift of $\sim 6.3$ based on recent EoR models \citep[e.g.,][]{Ishigaki2018, Naidu2020, Lidz2021}.
When using a shell implementation of the \citet{MiraldaEscude1998} model, we found that regions below $z\lesssim 6$ were too far away to have a discernible effect on the damping wing (see Section \ref{sec:shells}).
Since the sightline of GRB\,210905A would only have to be clear out to a redshift of $z\sim 6$ to get a non-detection, we run the simulation for a spherical region of radius $R = 125~\text{Mpc}\;h^{-1}$, the comoving distance between $z= 6.3$ and $z=6$.

We assume spherical neutral islands with $x_{\rm H_I} = 1$, and $x_{\rm H_I} = 0$ everywhere else, as is sometimes assumed for toy models of neutral gas distributions \citep[e.g.,][]{Spina2024}.
We determine the radius of each neutral island by pulling from a distribution based on the corresponding lengths of Ly$\beta$ gaps from \citet{Zhu2024}.
These neutral islands are then given a random position that does not overlap with any other neutral regions, or the GRB host galaxy at the center of the simulation.
The neutral islands continue to populate the simulation until their combined volume occupies at least 25\% of the simulation space.
10,000 random sightlines from the simulation are examined and used to estimate the percentage of over-ionized sightlines.
This process is re-run 10 times, and the percentage of over-ionized sightlines is averaged, to ensure that the result is not dictated by an unusual distribution of neutral islands.

We first run this simulation with the constraint that the center of all neutral islands must be more than $30~\text{Mpc}\;h^{-1}$ to account for the expected ionized zone around the GRB host galaxy at $x_{\rm H_I} \sim 0.25$ \citep{Lidz2021}.
This configuration results in $\sim 11.75\%$ of all sightlines being over-ionized.
While the $30~\text{Mpc}\;h^{-1}$ ionized region may be more realistic, we also perform the simulation without it to provide a more conservative estimate of $\sim 8.85\%$ of sightlines being over-ionized. 
If 8.85\% (roughly 1 in 12) to 11.75\% (roughly 1 in 9)  of sightlines at $z \sim 6.3$ are over-ionized, then there is a reasonable chance that the sightline of GRB\,210905A is over-ionized, and not representative of global neutral fraction at $z\sim 6.3$.

Given the simplicity of the simulation, there are multiple assumptions that may lead to over- or under-estimates of the fraction of ionized sightlines.
For one, the assumptions to make the neutral islands fully spherical, and to make them either fully ionized or fully neutral, reduces how broadly the neutral hydrogen is distributed in the simulations, which could potentially lead to an over-estimate.
There is also a limited amount of information about neutral island size distributions at redshifts greater than 6. 
The neutral island size distribution used in this simulation is largely based on dark troughs in Ly$\alpha$ transmission between $z\sim 5.5-6$ \citep{Zhu2024}.
Since lower redshifts are more ionized, we may be under-estimating the size of the neutral islands at $z\sim 6.3$.
However, even a small amount of neutral hydrogen ($x_{\rm H_I} \sim 10^{-4} - 10^{-3}$) can create or contribute to dark gaps \citep{Zhu2022, Gaikwad2023}, so the full length of the dark troughs may be not be exclusively caused by highly neutral regions.
If residual neutral hydrogen in highly ionized regions is also contributing to the lengths of the dark troughs from \citet{Zhu2024}, then we may instead be overestimating the radii of the neutral islands.
The simulation also takes sightlines with any interaction with a neutral island, regardless of how short the distance, to be non-ionized sightlines. 
Short segments of neutral hydrogen, especially ones farther away from the GRB redshift, may not have a strong enough impact on the damping wing to be detectable, and could still appear to be an overly-ionized sightline in the context of a damping wing analysis, even though it is not fully ionized.
In this way, the simulation may also be under-estimating the fraction of over-ionized sightlines.

In comparison, the mean free path $\lambda_{\rm mfp}$ of photons through the IGM has been estimated using Ly$\alpha$ and Ly$\beta$ transmission profiles, with estimates of $\lambda_{\rm mfp}=0.75^{+0.65}_{-0.45}$ proper Mpc (pMpc) at $z = 6$ \citep[or $\sim 3.54~\text{Mpc}\;h^{-1}$;][]{Becker2021} and $\lambda_{\rm mfp} = 0.81^{+0.73}_{-0.48}$ pMpc at $z = 5.93$ \citep[or $3.78\;\text{Mpc\;}h^{-1}$;][]{Zhu2023}.
While no direct measurements of the mean free path have been made beyond $z\sim 6$, hydrodynamical simulations suggest a mean free path of $\sim 0.2$~pMpc \citep[$\sim 1~\text{Mpc}\;h^{-1}$;][]{Lewis2022}.
For any of these mean free paths, the probability of a photon reaching $125~\text{Mpc}\;h^{-1}$ is virtually 0, which is not consistent with the findings of our simulation.
One cause of this discrepancy is that the mean free path is affected by small amounts of neutral hydrogen in highly ionized regions, so it is not exclusively dependent on interactions with mainly neutral regions.
If the mean free path at $z\sim 6$ is mainly limited by residual neutral hydrogen in ionized regions, then many photons traveling along over-ionized sightlines would still likely be absorbed, and the fraction of photons reaching $125~\text{Mpc}\;h^{-1}$ would be much lower than the fraction of over-ionized sightlines.
Another potential contributer to the inconsistency is large-scale fluctuations in the ionizing UV background.
Both neutral islands and large-scale fluctuations in the UV background can cause dark gaps in transmission that would impact the optical depth, but the latter would have no impact on a damping wing \citep{Zhu2023, Zhu2024}.
This could explain why the probability for an over-ionized sightline is so much higher than the probability for a photon to be un-attenuated between $z\sim 6-6.3$.
However, this large discrepancy still highlights the fact that our simulation is idealized, and may be over-estimating the number of clear sightlines at $z\sim 6.3$.

\subsection{Increasing the Sample of High-Redshift GRBs}
\label{sec:highzgrbs}
There have been a few other GRB damping wing analyses to measure the fraction of neutral hydrogen in the IGM. In an analysis of GRB\,050904, another $z\sim 6.3$ GRB, \citet{Totani2006} found a best fit neutral fraction of 0 with upper limits of $<0.17$ and $<0.60$ at the 1- and 2-$\sigma$ confidence levels, respectively.
A similar analysis was performed by \citet{Patel2010} on GRB\,080913 at $z\sim 6.7$, but they were unable to constrain $x_{\rm H_I}$. 
Multiple analyses were performed on the red damping wing of GRB\,130606A at $z\sim 5.9$, resulting in varying results. 
An analysis using data from the MMT Blue Channel Spectrograph \citep{Schmidt1989} and the Gemini Multi-Object Spectrograph \citep[GMOS;][]{Hook2004} on Gemini-North resulted in a 2-$\sigma$ upper limit of $x_{\rm H_I} < 0.11$ \citep{Chornock2013}, while damping wing analysis of data from the Faint Object Camera and Spectrograph \citep[FOCAS;][]{Kashikawa2002} on Subaru resulted in $x_{\rm H_I} \sim 0.1 - 0.5$ depending on model assumptions \citep{Totani2014}. 
Two teams using data from X-Shooter found a 3-$\sigma$ upper limit of $x_{\rm H_I} < 0.05$ \citep{Hartoog2015} and a detection of $x_{\rm H_I} = 0.061\pm 0.007$ \citep{Totani2016}. 
Finally, \citet{Melandri2015} examined the red damping wing of GRB\,140515A at a redshift of $z \sim 6.33$, and found an upper limit of $x_{\rm H_I} \lesssim 0.002$. 
Future analyses of GRB Ly$\alpha$ damping wings will require a careful treatment of the data. It will also be important to increase the number of high-redshift GRBs with high quality spectra to perform more robust measurements of $x_{\rm H_I}$.

Increasing the sample of high-redshift GRBs will impact our understanding of the EoR, as the more neutral fraction estimates we can obtain at a given redshift, the more confidence we can have in our understanding of the ionization state as a function of redshift.
Future missions like the \emph{Space Variable Objects Monitor} \citep[\emph{SVOM};][]{Wei2016, Atteia2022}, as well as proposed missions like the \emph{Gamow Explorer} \citep[\emph{Gamow};][]{White2021} and  \emph{Transient High-Energy Sky and Early Universe Surveyor} \citep[\emph{THESEUS};][]{Amati2021}, will be crucial to find a large sample of high-$z$ GRBs.
New instruments and facilities, such as SCORPIO \citep{SCORPIO} on Gemini and a new generation of 30-meter telescopes \citep{Neichel2018}, with spectroscopy capabilities over a wide wavelength range, will be important for detailed analyses along the lines of what we present in this paper.

\section{Conclusions}

GRBs are excellent probes of the high-redshift Universe.
They are useful for studying the EoR, as they can be used to track its progression by measuring the fraction of neutral hydrogen in the IGM at different redshifts.
GRB\,210905A was a $z \sim 6.3$ burst that had one of the most luminous late-time optical afterglows ever observed \citep{Rossi2022}, and was used to probe the chemical composition of its host galaxy and multiple intervening absorbing systems \citep{Saccardi2023}.
We have presented a detailed analysis of the afterglow spectrum of GRB\,210905A to estimate the fraction of neutral hydrogen in the IGM surrounding the its host galaxy.
Based on the preferred model with two uncoupled DLAs, we find a 3-$\sigma$ upper limits on the neutral fraction of $x_{\rm H_I} \lesssim 0.15$ and $x_{\rm H_I} \lesssim 0.23$ for the \citet{MiraldaEscude1998} and \citet{McQuinn2008} models, respectively, which indicates that the IGM around the GRB host galaxy is highly ionized. We note that performing fits with a coupling between the DLAs, based on information from the high-quality spectra, leads to tighter upper limits of $x_{\rm H_I} \lesssim 0.04$.
This result is generally in agreement with other neutral fraction measurements performed with other high-redshift probes of the IGM.
We discussed the complications of this particular analysis brought about by possible curvature in GRB afterglow spectra and telluric lines. We also consider avenues for pushing forward this field of study.
More high-redshift GRBs will be crucial to the continued study of the EoR and its progress at different points in cosmological history.

\section*{Acknowledgements}

We thank the referee for their timely and constructive feedback. Based on observations collected at the European Southern Observatory under ESO programme 106.21T6.014.
This work made use of data supplied by the UK Swift Science Data Centre at the University of Leicester.
A.R. acknowledge support from PRIN- MIUR 2017 (grant 20179ZF5KS).

\section*{Data Availability}

The data presented in this paper are available upon request to the first author.



\bibliographystyle{mnras}
\bibliography{Bib} 




\appendix

\section{Additional Ionized Bubble Model Fits}
\label{sec:additionalMcquinn}

\begin{figure}
    \centering
    \includegraphics[width = \linewidth,trim={0 110 0 150},clip]{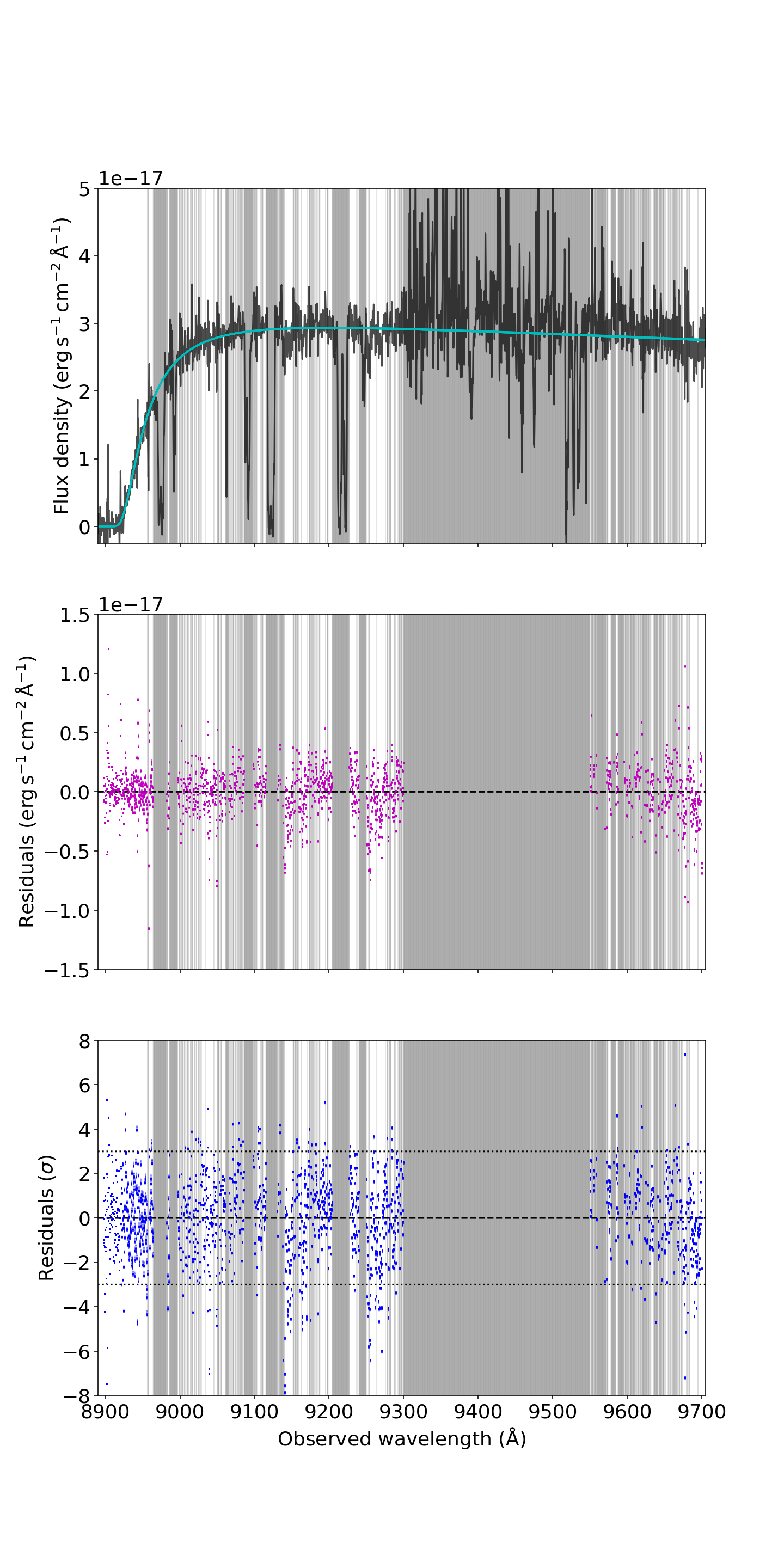}
    \caption{Fit using the \citet{McQuinn2008} model with uncoupled column densities for the two DLAs, a fixed spectral index of $\beta = 0.40$, and an $R_{\rm b}$ upper bound of $130~\text{Mpc}\;h^{-1}$ or $193~\text{Mpc}$. See Figure \ref{fig:coupled} for description of the three panels.}
    \label{fig:mcquinnplot130}
\end{figure}

\begin{figure}
    \centering
    \includegraphics[width = \linewidth,trim={0 0 28 10},clip]{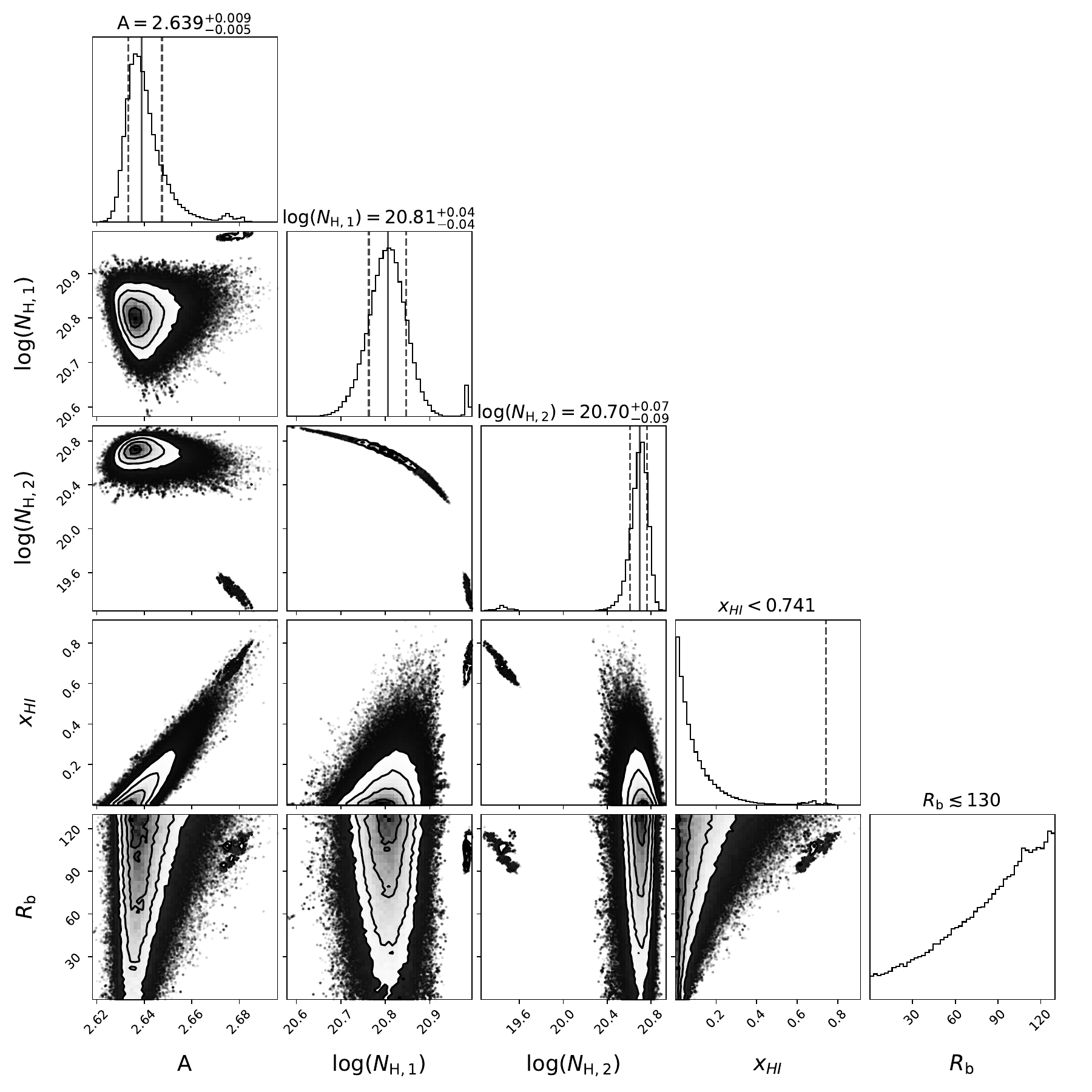}
    \caption{Corner plot for the fit shown in Figure \ref{fig:mcquinnplot130}. The solid grey lines represent the 50\% centile, while the dashed grey lines show the 1-$\sigma$ error ranges (16\% and 84\% centiles). The normalization is given in $10^{-17}\,\ergscmA$ and the ionized bubble radius is given in Mpc. $N_{\rm H,1}$ corresponds to the DLA at $z = 6.3186$, and $N_{\rm H,2}$ corresponds to the DLA at $z = 6.3118$.}
    \label{fig:mcquinn_corner130}
\end{figure}

\begin{figure}
    \centering
    \includegraphics[width = \linewidth,trim={0 110 0 150},clip]{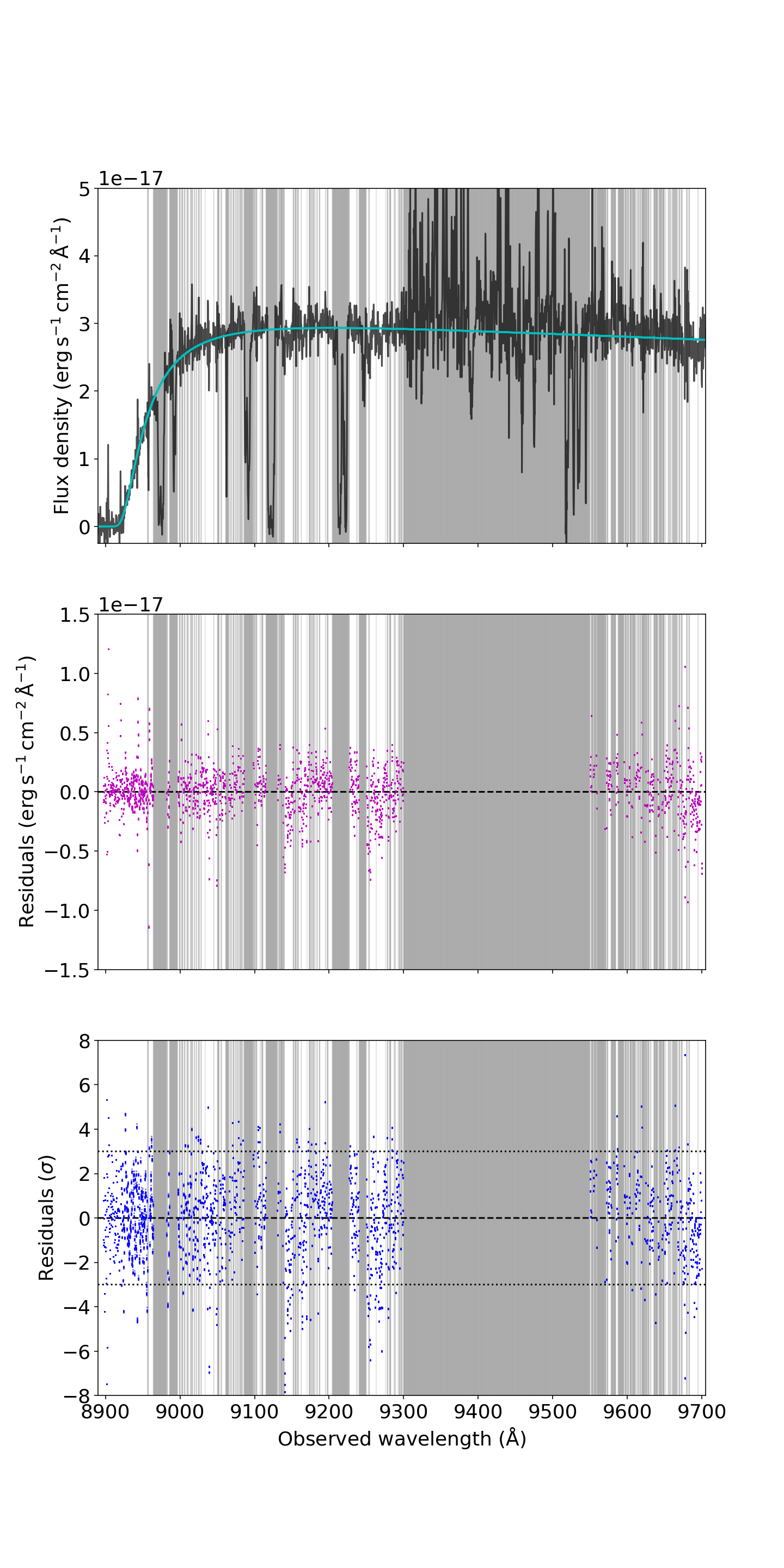}
    \caption{Fit using the \citet{McQuinn2008} model with uncoupled column densities for the two DLAs, a fixed spectral index of $\beta = 0.40$, and an $R_{\rm b}$ upper bound of $355~\text{Mpc}\;h^{-1}$ or $527~\text{Mpc}$. See Figure \ref{fig:coupled} for description of the three panels.}
    \label{fig:mcquinnplot355}
\end{figure}

\begin{figure}
    \centering
    \includegraphics[width = \linewidth,trim={0 0 28 10},clip]{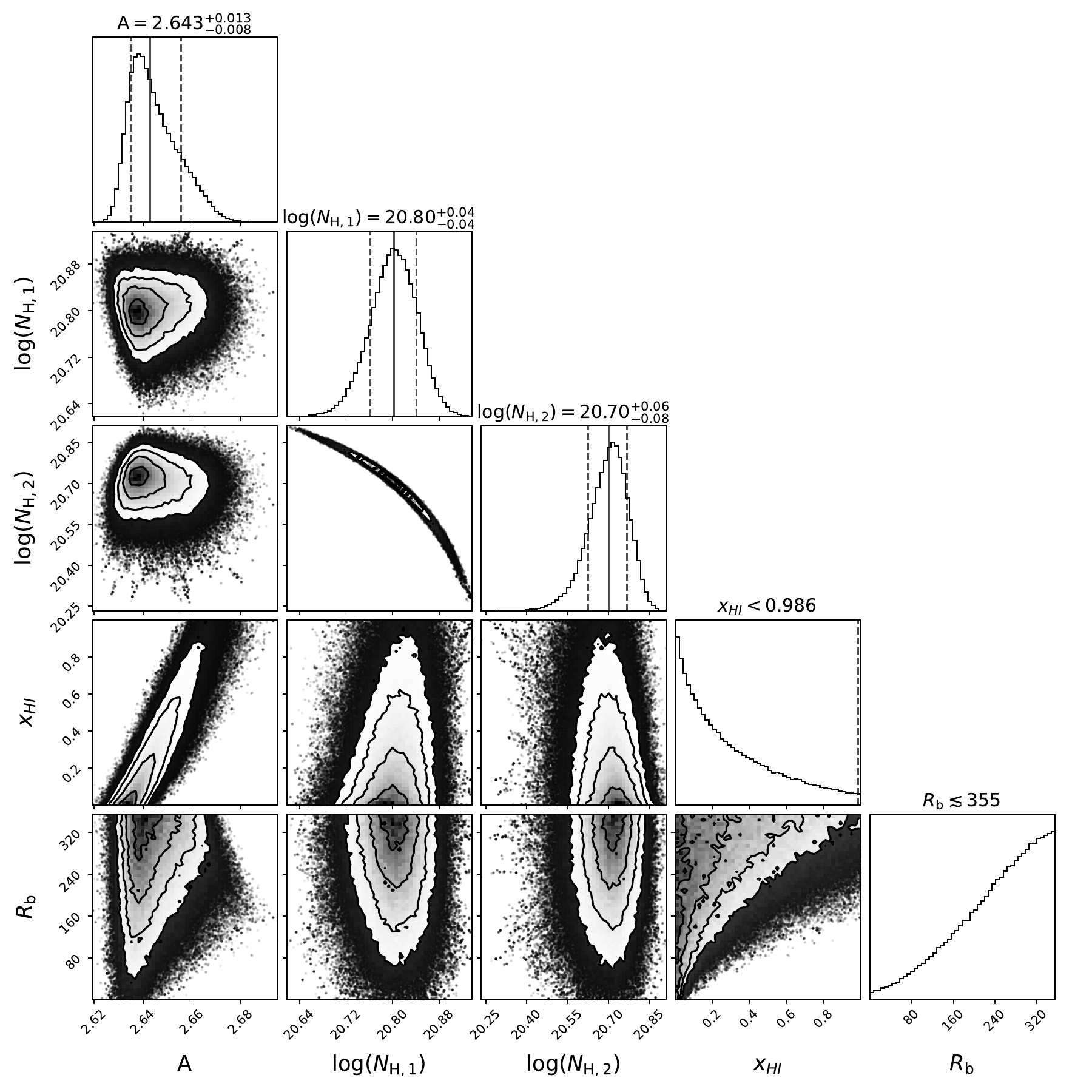}
    \caption{Corner plot for the fit shown in Figure \ref{fig:mcquinnplot355}. The solid grey lines represent the 50\% centile, while the dashed grey lines show the 1-$\sigma$ error ranges (16\% and 84\% centiles). The normalization is given in $10^{-17}\,\ergscmA$ and the ionized bubble radius is given in Mpc. $N_{\rm H,1}$ corresponds to the DLA at $z = 6.3186$, and $N_{\rm H,2}$ corresponds to the DLA at $z = 6.3118$.}
    \label{fig:mcquinn_corner355}
\end{figure}

In Section \ref{sec:McQuinn}, we show that the size of the ionized bubble around the GRB, $R_{\rm b}$, is not well constrained and tends towards the largest possible bound allowed by the prior. We also performed fits with $R_{\rm b}$ upper bounds of $R_{\rm b} < 130~\text{Mpc}$ (corresponding to $z \sim 6.0$; see Figures \ref{fig:mcquinnplot130} and \ref{fig:mcquinn_corner130}) and $R_{\rm b} < 355~\text{Mpc}$ (corresponding to $z \sim 5.5$; see Figures \ref{fig:mcquinnplot355} and \ref{fig:mcquinn_corner355}), to ensure that this behavior is not specific to the chosen prior. We find that the $R_{\rm b}$ posterior is always unconstrained but tends towards the largest possible $R_{\rm b}$ regardless of our choice of upper limit. This indicates that the shape of the $R_{\rm b}$ posterior is not a product of our choice of bounds for the $R_{\rm b}$ prior.

As $R_{\rm b}$ increase, the upper limit on $x_{\rm H_I}$ also increases, but this is expected because the farther neutral hydrogen becomes from the GRB, the less it impacts the shape of the damping wing and the harder it becomes to constrain its value. This behavior can also be seen in the $x_{\rm H_{I}}$ vs. $R_{\rm b}$ corner plot panels (see Figures \ref{fig:mcquinn_corner130} and \ref{fig:mcquinn_corner355}), as there is a wider spread of neutral fraction values at higher values of $R_{\rm b}$. In all cases the posterior is still most densely populated around 0.

\section{Shell implementation}

\begin{figure}
    \centering
    \includegraphics[width = \linewidth, trim={0, 120, 0, 140}, clip]{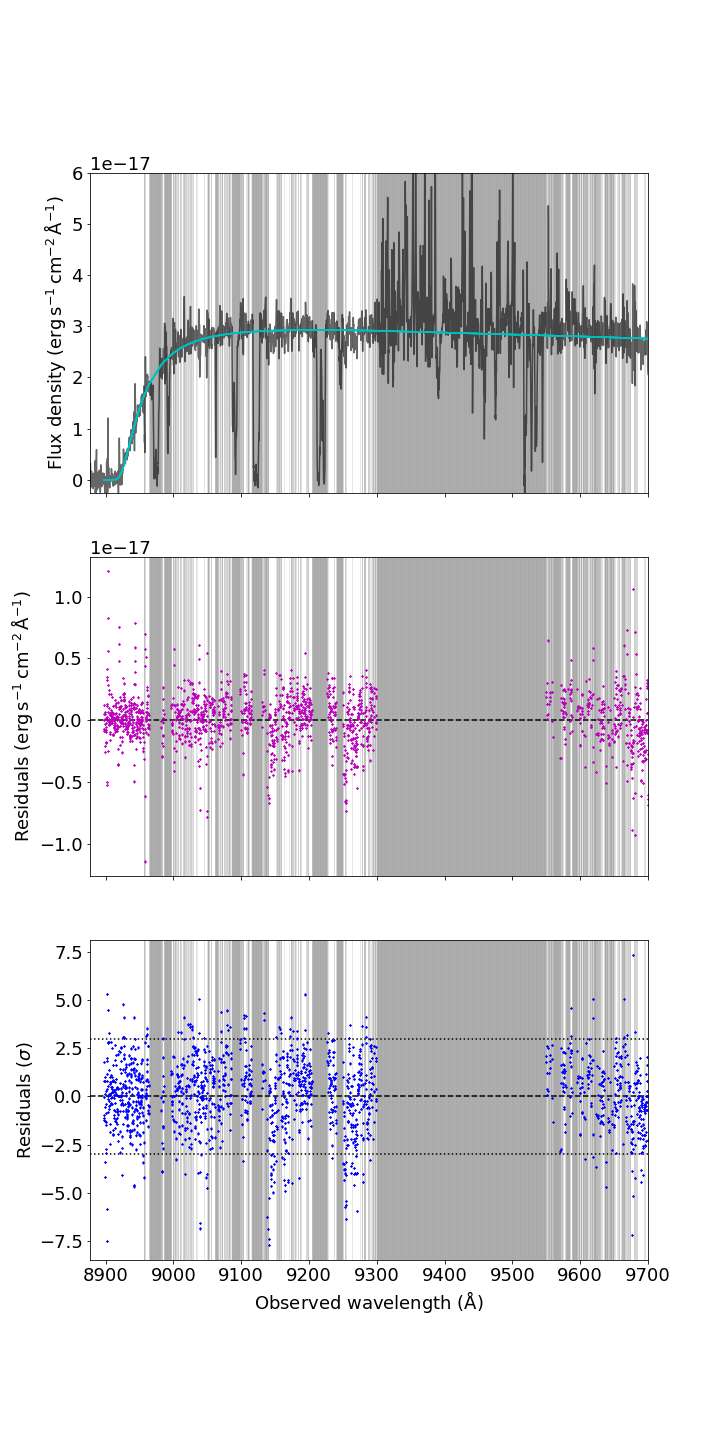}
    \caption{Fit using the shells implementation of the \citet{MiraldaEscude1998} model with uncoupled column densities for the two DLAs, a fixed spectral index of $\beta = 0.40$, and shells of width $\Delta z = 0.1$ out to $z = 5.5$. See Figure \ref{fig:coupled} for description of the three panels.}
    \label{fig:shellFit}
\end{figure}

\begin{figure*}
    \centering
    \includegraphics[width = \linewidth]{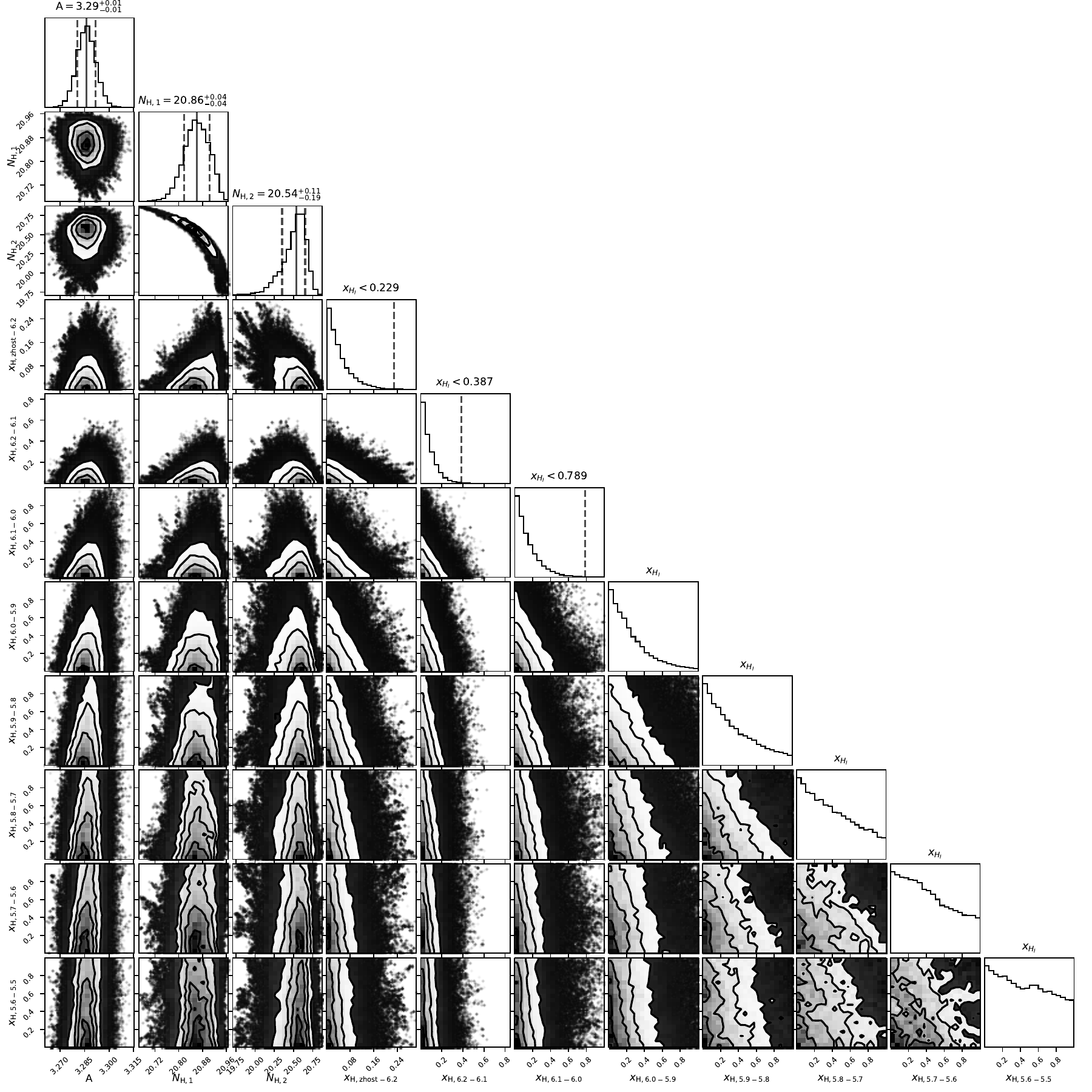}
    \caption{Corner plot associated with Figure \ref{fig:shellFit}. The upper limit on $x_{\rm H_I}$ gradually increase with redshift. at redshifts sufficiently far from the GRB redshift ($z \sim 5.7-5.5$), the posterior for $x_{\rm H_I}$ flattens out, and any impact of these regions on the shape of the Ly$\alpha$ damping wing can no longer be distinguished.}
    \label{fig:shellCorner}
\end{figure*}

To better account for clumpiness in the IGM, we also implement the \citet{MiraldaEscude1998} model in shells of width $\Delta z = 0.1$ from the GRB redshift out to $z = 5.5$ (see Section \ref{sec:shells}). The fit and corner plot for this implementation are shown in Figures \ref{fig:shellFit} and \ref{fig:shellCorner}, respectively.

\bsp	
\label{lastpage}
\end{document}